\documentclass[12pt]{iopart}
\usepackage{iopams}
\usepackage{graphicx}
\usepackage{cite}

 \eqnobysec

\begin{document}

\newcommand{\dd}{\mathrm{d}}
\newcommand{\ii}{\mathrm{i}}
\newcommand{\ee}{\mathrm{e}}
\newcommand{\Sp}{\mathrm{Sp}}
\newcommand{\heav}{\theta}
\newcommand{\dirac}{\delta}
\newcommand{\kronek}{\delta}
\newcommand{\lindhard}{\mathrm{L}}
\newcommand{\jell}{\mathrm{jell}}
\newcommand{\unif}{\mathrm{unif}}
\newcommand{\bulk}{\mathrm{bulk}}
\newcommand{\surf}{\mathrm{surf}}
\newcommand{\const}{\mathrm{const}}

\newcommand{\arctg}{\mathrm{arctg}}

\title[Semi-infinite jellium: thermodynamic potential, chemical potential, surface energy]
      {Semi-infinite jellium: thermodynamic potential, chemical potential, surface energy}

\author{P P Kostrobij, B M Markovych}

\address{Lviv Polytechnic National University, 12 Bandera Str., 79013 Lviv, Ukraine}
\ead{bogdan\_markovych@yahoo.com}
\vspace{10pt}
\begin{indented}
\item[]January 2015
\end{indented}

\begin{abstract}
 General expression for the thermodynamic potential of the
 model of semi-infinite jellium is obtained.
 By using this expression, the surface energy for infinite barrier model is calculated.
 The behavior of the surface energy and of chemical potential as functions of the Wigner-Seitz radius
 and the influence of the Coulomb interaction between electrons on the calculated values
 is studied.
 It is shown that taking into account the Coulomb interaction between electrons leads to growth of the surface energy.
 The surface energy is positive in the entire area of the Wigner-Seitz radius.
 It is shown that taking into account the Coulomb interaction between electrons leads to a decrease of the chemical potential.
\end{abstract}


\vspace{2pc}

%
%
%

 \section{Introduction}

 The development of quantum-statistical theory of Fermi systems with interfaces
 is one of the most important problems of contemporary statistical physics.
 In particular, the richness of surface phenomena and
 the rapid development of experimental methods of investigation of surfaces requires
 the development of theory of such systems.

 The most popular theoretical method for studying in the area of research is
 the density functional theory~\cite{Dreizler,Lundqvist,Partenskii},
 that have been developed from the well-known Thomas-Fermi method for atoms.
 By construction, the density functional theory is the one-particle approach and can not properly take into account
 the many-body correlation effects.
 Therefore, the energy functionals for inhomogeneous systems are mostly used in the local density approximation~\cite{Lundqvist},
 namely the electron density distribution $n(\mathbf{r})$ is substituted by the average electron density $n=\const$
 in  the well-known expressions of the theory for homogeneous systems.
 This approach is questionable \cite{Sarry},
 since the presence of the interface brings both quantitative and qualitative changes of various characteristics
 of an electronic system, e.g. the image forces,
 that cannot be obtained from the density functional theory in principle.

 Density functional theory has a characteristic problem of surface energy.
 Namely, the surface energy of semi-infinite jellium  calculated in this theory turns out to be negative for large values of the electron concentration (${r_\mathrm{s}<2.5\,a_\mathrm{B}}$, where ${r_\mathrm{s}}$ is the Wigner-Seitz radius)~\cite{Lang}.
 This is physically incorrect.
 The surface energy must be positive, otherwise the metal would spontaneously split.
 At present, general belief is that the cause of negativity of the surface energy is
 the replacement of the discrete ionic lattice by uniform positive background.
 Thus, in the work of Lang and Kohn~\cite{Lang} a discrete lattice is accounted for
 using the first-order perturbation theory in the pseudopotential.
 As a result,
 the surface energy becomes positive and is satisfactorily consistent with the experimental data for a number of simple metals~\cite{Lang}.
 A variational procedure has been developed by Monnier and Perdew~\cite{Monnier,Perdew}
 to take into account the averaged effect with the introduction of discrete additions to the potential inside a metal,
 which depends on the structure of a lattice and the surface.
 The obtained results for the surface energy were very close to
 the results of Lang and Kohn and were in better agreement with the experimental data.
 Later, Appelbaum and Hamann \cite{Appelbaum}
 used the local density approximation and performed the calculations for the Cu(111) surface,
 with the account of the discrete ionic lattice without perturbation theory.
 They have obtained good agreement with the experimental data.
 The authors of all these studies have assumed that the non-local exchange-correlation effects are negligible and can be omitted.
 In the works \cite{Paash4,Paash5},
 the calculations of the surface energy were performed with varying position of the last (exposed) ion layer
 and good agreement with experimental data for the surface energy of simple metals was obtained.
 In the works \cite{Perdew3,Shore,Rose2}, a stabilized jellium model has been proposed.
 In this model the pseudopotential correction, which is averaged over the Wigner-Seitz cell,
 is incorporated into the effective potential inside the metal.
 This model yields positive values for the surface energy.
 Thus, the consideration of discrete ionic lattice permits to solve the problem of negative values of the surface energy.

 However, it is not clear that neglecting the discreteness is the principal and only reason of discrepancy
 between the theory and experiment.
 Probably the theory could be improved, still remaining in the framework of the jellium model.
 There have been attempts to go beyond the local density approximation.
 Namely, Schmit and Lucas \cite{Schmit}, Craig \cite{Craig} and Peuckert \cite {Peuckert} have considered
 certain non-local contributions to the exchange-correlation term to the surface energy,
 due to the change of zero energy of plasmons and the appearance of surface modes during separation of crystal into fragments.
 From a good agreement of this contribution with the experimental data for the surface energy,
 the authors of these works proposed to identify this contribution with the total surface energy,
 suggesting that other contributions, that have not been accounted for, canceled each other.
 This approach is actively debated in Refs. \cite{Kohn,Feibelman,Jonson,Craig2,Heinrichs,Harris,Harris2,Paash2}.
 The expressions for the exchange-correlation energy of bounded electron gas have been obtained by
 Harris and Jones \cite{Harris,Harris2}, Wikborg and Inglesfield \cite{Wikborg}, Johnson and Srinivasan \cite{Jonson2},
 focusing in the analysis of non-local effects.
 The exchange part of the surface energy was calculated for electrons in a potential box with infinitely high walls~\cite{Harris,Harris2},
 and the exchange-correlation and exchange parts were calculated in the random phase approximation~\cite{Wikborg}.
 A comparison of these results with those calculated in the local density approximation
 has showed that the exchange-correlation parts differ by about 10\%,
 the exchange parts differ by 50\%, and the correlation parts differ by 6~times.
 The local density approximation works much better for the sum of the exchange and correlation parts of energy
 rather than for each individual contribution (see also \cite{Lang2}).
 Further calculations, using more realistic models of the surface barrier~\cite{Paash3,Mamoun,Wang,Sahni,Sahni2},
 namely the gradient expansions~\cite{Wang,Rasolt,Rasolt2,Rose,Gupta,Perdew2,Langreth,Langreth2},
 the analysis of Langreth and Perdew \cite{Langreth,Langreth2},
 have found a decisive contribution of the local density approximation into the exchange-correlation part of energy as well.
 In particular, the calculations~\cite{Rose} show that the non-local corrections do not exceed 16\%,
 although the relative contribution of non-locality into the total surface energy can be much larger (up to 40\%)
 because for many metals the exchange-correlation part of the surface energy is greater than the total surface energy \cite{Lang}.

 More recent studies have renewed the debate about the correctness of the application of the local density approximation
 in the calculation of the surface energy.
 Using the Fermi hypernetted-chain equations~\cite{Krotscheck,Krotscheck2} have obtained significantly higher values of the surface energy than calculated by Lang and Kohn in the local density approximation~\cite{Lang}.
 In contrast, the values of the surface energy obtained in the calculations using density functional theory \cite{Zhang} with nonlocal functional of Langreth and Mehl \cite{Langreth3}, are much closer to the results obtained in the local density approximation.
 Calculations of the surface energy by using quantum Monte Carlo method \cite{Li,Acioli}  have shown that the values of the surface energy obtained at high concentrations (${r_\mathrm{s}\leqslant2.07\,a_\mathrm{B}}$) are in a good agreement with the results obtained by using nonlocal functional, but at lower concentrations ${r_\mathrm{s}\geqslant3.25\,a_\mathrm {B}}$) they are in a good agreement with the results of Krotscheck and Kohn \cite{Krotscheck,Krotscheck2}.
 On the other hand, Pitarke has concluded that the local density approximation leads to a small error in the exchange-correlation of the surface energy \cite{Pitarke}.
 In this work, the long-range correlations are taken into account self-consistently in the random phase approximation, while the short-range correlations are included in the time-dependent local density approximation \cite{Runge,Gross}.

 In Ref. \cite{UFZ2002} the surface energy is calculated by using the one- and two-particle distribution functions of electrons,
 that are obtained in \cite{JPS2003_2}
 taking into account the Coulomb interaction between electrons.
 It is found that the surface energy is positive in the entire region of electron concentration.

 Takahashi and Onzawa have calculated the electron density distribution and the surface energy of non-interacting electron gas for finite barrier model \cite{Takahashi}.
 It is interesting that the electron density without self-consistency is very close to the self-consistent calculations of Lang and Kohn \cite{Lang}, and the surface energy is positive for all concentrations of electrons.
 Moreover their result is close to the result of Lang and Kohn at low concentrations.

 In the present work, our principle objective is to construct a consistent quantum-statistical theory of a simple metal with the interface `metal--vacuum' in the framework of the jellium model.
 An effective potential of inter-electron interaction,
 which we have studied recently in \cite{CMP2003,JPS2003_1,CMP2006,UFZ2007,FMMIT2013},
 is crucial for calculating of the general expression for thermodynamic potential.
 It is shown that within certain approximations the thermodynamic potential
 can be represented as a functional of the one- and two-particle distribution functions of electrons.
 At low temperatures, the nonlinear algebraic equation for the chemical potential and a general expression for the internal energy are obtained from the thermodynamic potential.
 It is shown that obtained equation for the chemical potential of non-homogeneous system is
 similar to the equation for a homogeneous system.
 The chemical potential is calculated as a function of the Wigner-Seitz radius.
 The one- and two-particle distribution function of electrons are calculated for the infinite barrier model.
 The expressions for the extensive and surface contributions to the internal energy are obtained.
 For the same model, the calculation of surface contribution to the internal energy,
 which is the surface energy at low temperature, is performed.
 The behaviour of the surface energy as a function of the Wigner-Seitz radius is studied.
 We have found that the surface energy calculated in the present work is positive in the entire concentration range
 typical for metals, and at low concentrations is consistent with the calculations of Lang and Kohn~\cite{Lang}.
 The influence of the Coulomb interaction between electrons on the calculated characteristics are studied.
 Detailed calculations of these properties is given in Ref.~\cite{1402U}.

 \section{Model}

 We consider a system of $N$ electrons in the volume $V=SL$
 in the field of positive charge with the distribution
 \begin{equation}\label{PositiveCharge}
   \varrho_\jell({\bf R}_{||},Z)\equiv\varrho_\jell(Z)=\varrho_0\heav(-Z)=
    \left\{\begin{array}{ll}
                      \varrho_0, & Z\leqslant0 \\
                      0, & Z>0
                    \end{array}
    \right.    ,
 \end{equation}
 where $Z=0$ is the dividing plane,
 $\heav(x)$ is the Heaviside step function,
 ${\bf R}_{||}=(X,Y)$,
 $X,Y\in [-\sqrt{S}/2,+\sqrt{S}/2]$,
 $Z\in\left[-L/2,+L/2\right]$.
 The condition of electroneutrality is satisfied,
 \begin{equation}\label{2.1}
  \lim_{S,L\to\infty}
  \int\limits_S \! \dd{\bf R}_{||} \!\!
  \int\limits_{-L/2}^{+L/2} \!\!\!\! \dd Z \,
  \varrho_\jell({\bf R}_{||},Z) = e N, \; e>0,
 \end{equation}
 moreover in the thermodynamic limit we have,
 \begin{equation}\label{ThermodLimit}
    \lim_{N,S,L\to\infty}\frac{eN}{SL}=
    \lim_{N,V\to\infty}\frac{eN}{V/2}=
    \varrho_0.
 \end{equation}

 This model system is known as ``semi-infinite jellium''
 and it is one of the simplest models of semi-infinite metal,
 which satisfactorily describes simple metals.
 The Hamiltonian of the model is,
\begin{eqnarray}\label{Hamiltonian}
    H_\jell &=  -\frac{\hbar^2}{2m}\sum\limits_{i=1}^N\Delta_i
        + \frac12\sum\limits_{i\ne j=1}^N\frac{e^2}{|{\bf r}_i-{\bf r}_j|}
        - \sum\limits_{j=1}^N\int\limits_V \! \dd{\bf R}\,\frac{e\varrho_\jell(\mathbf{R})}{|\mathbf{r}_j-\mathbf{R}|}\nonumber\\
        &\quad+\frac12\int\limits_V \! \dd{\bf R}_1 \!\! \int\limits_V \! \dd{\bf R}_2 \,
         \frac{\varrho_\jell({\bf R}_1)\varrho_\jell({\bf R}_2)}{|{\bf R}_1-{\bf R}_2|},
\end{eqnarray}
 where $\mathbf{r}_j$ is the position of $j$-th electron;
 the first term is the kinetic energy of electrons
 ($m$ is the electron mass),
 the second term is the potential energy of the inter-electron interaction,
 the third term is energy of interaction of electrons with the positive charge,
 the fourth term is potential energy of the positive charge.

 From the Hamiltonian (\ref{Hamiltonian})
 we extract a Hamiltonian of the infinite jellium model~$H_\jell^\unif$
\begin{eqnarray}\label{HamiltonianUnif}
    H_\jell^\unif &=  -\frac{\hbar^2}{2m}\sum\limits_{i=1}^N\Delta_i
        + \frac12\sum\limits_{i\ne j=1}^N\frac{e^2}{|{\bf r}_i-{\bf r}_j|} \nonumber
        - \sum\limits_{j=1}^N\int\limits_V \! \dd{\bf R}\,\frac{e^2N/V}{|\mathbf{r}_j-\mathbf{R}|}\\
        &\quad+ \frac12\int\limits_V \! \dd{\bf R}_1 \!\! \int\limits_V \! \dd{\bf R}_2 \,
         \frac{(eN/V)^2}{|{\bf R}_1-{\bf R}_2|},
\end{eqnarray}
 here physical meaning of the terms are similar to the terms of the Hamiltonian~(\ref{Hamiltonian}).

 Thus we get,
\begin{equation}\label{Hamiltonian2}
   \fl H_\jell = H_\jell^\unif
        + \sum\limits_{j=1}^NV_\surf(\mathbf{r}_j) + \frac12\int\limits_V \! \dd{\bf R}_1 \!\! \int\limits_V \! \dd{\bf R}_2 \,
         \frac{\varrho_\jell({\bf R}_1)\varrho_\jell({\bf R}_2)-(eN/V)^2}{|{\bf R}_1-{\bf R}_2|} ,
\end{equation}
 where
 \begin{equation}\label{SurfacePotential}
    V_\surf(\mathbf{r}_j)=\int\limits_V \! \dd{\bf R}\,
        \frac{e\big(eN/V-\varrho_\jell(\mathbf{R})\big)}{|\mathbf{r}_j-\mathbf{R}|}
 \end{equation}
 is the surface potential acting on the electron.
 This potential is formed by the deviation of the positive charge distribution from the uniform one.
 So, if instead of $\varrho_\jell({\bf R})$ we put the uniform distribution $eN/V$,
 then the surface potential and the last term in the Hamiltonian (\ref{HamiltonianUnif})
 disappear and one obtains,
 \[
  \lim_{\varrho_\jell\to \frac{eN}{V}}H_\jell=H_\jell^\unif.
 \]

 It should be noted that
 as a consequence of the symmetry of the model,
 the surface potential $V_\surf(\mathbf{r})$
 is a function of the normal to the dividing plane coordinates of the electron only,
 the motion of the electron in a plane parallel to the dividing plane is free, i.e.,
 \[
  V_\surf(\mathbf{r})\equiv V_\surf(z).
 \]

 In order to calculate the thermodynamic potential of the system, it is convenient to
 present the Hamiltonian (\ref{Hamiltonian2}) in the secondary quantization representation.

 \section{Secondary quantization representation}

 We introduce the single-particle wave functions $\Psi_a({\bf r})$ and the corresponding
 energies $E_a$ of the electron in the field of the surface potential $V_\surf(z)$,
 \begin{equation}\label{3.1}
 \left[-\frac{\hbar^2}{2m}\Delta+V_\surf(z)\right]
 \Psi_a({\bf r})=E_a\Psi_a({\bf r}),
 \end{equation}
 which we use to construct the representation of the secondary quantization.

 Since the potential in the stationary Schr\"odinger equation (\ref{3.1})
 depends only on the normal to the dividing plane coordinate of the electron,
 the variables can be separated.
 Then we obtain,
 \begin{equation}\label{3.2}
 E_a=\frac{\hbar^2p^2}{2m}+\varepsilon_\alpha,\;a=({\bf
 p},\alpha),
 \end{equation}
 \begin{equation}\label{3.3}
 \Psi_a({\bf r})=\frac1{\sqrt{S}} \,
 {\rm e}^{\mathrm{i}{\bf p}{\bf r}_{||}} \varphi_\alpha(z).
 \end{equation}
 where ${\bf r}_{||}$ is two-dimensional coordinate of the electron
 in the plane parallel to the dividing plane,
 $\hbar{\bf p}$ is the moment of the electron in this plane, and
 \begin{equation}\label{impulsP}
  {\bf p}=(p_x,p_y),\quad
    p_{x,y}=\frac{2\pi n_{x,y}}{\sqrt{S}},\quad
    n_{x,y}=0,\pm1,\pm2,\ldots,
 \end{equation}
 $\alpha$ is some quantum number,
 that depends on the form of the surface potential,
 the functions $\varphi_\alpha(z)$ satisfy the one-dimensional stationary Schr\"odinger equation,
 \[
   \left[-\frac{\hbar^2}{2m}\frac{\dd^2}{\dd z^2}+V_\surf(z)\right] \varphi_\alpha(z)=\varepsilon_\alpha\varphi_\alpha(z).
 \]

 In the secondary quantization representation constructed by means
 of the wave functions (\ref{3.3}), the Hamiltonian of the system becomes,
 \begin{equation}\label{3.10}
  \fl  H=\sum_{{\bf p},\alpha}E_\alpha({\bf p})
         a_\alpha^\dagger({\bf p})
         a_\alpha^{\vphantom{\dagger}}({\bf p})-
         \frac1{2S}N\sum_{\mathbf{q}\neq0}\nu({\bf q},0)
         +\frac1{2SL}\sum_{\mathbf{q}\neq0}\sum_k\nu_k({\bf q})\rho_k({\bf q})\rho_{-k}(-{\bf q}),
 \end{equation}
 where $a_{\alpha}^\dagger({\bf p})$,
 $a_{\alpha}^{\vphantom{\dagger}}({\bf p})$ are the operators of electron creation
 and annihilation, respectively, in the state $({\bf p},\alpha)$,
 and the standard commutation relations are,
 \begin{equation}\label{3.6}
 \left\{a_{\alpha_1}^{\vphantom{\dagger}}({\bf p}_1),
 a_{\alpha_2}^\dagger({\bf p}_2)\right\}=
 \delta_{{\bf p}_1,{\bf p}_2}\delta_{\alpha_1,\alpha_2},
 \end{equation}
 \begin{equation}\label{operN}
  N=\sum\limits_{{\bf p},\alpha}
    a_{\alpha}^\dagger({\bf p})
    a_{\alpha}^{\vphantom{\dagger}}({\bf p})
 \end{equation} is the particle number operator,
 $\nu\left({\bf q},0\right)=\frac{2\pi e^2}{q}$,
 $\nu_k({\bf q})=4\pi e^2/({\bf q}^2+k^2)$ is the Fourier-transform of the Coulomb interaction,
 $q_{x,y}=\frac{2\pi}{\sqrt{S}}m_{x,y}$,
 $m_{x,y}=0,\pm1,\pm2,\ldots$,
 $k=\frac{2\pi}{L}n$,
 $n=0,\pm1,\pm2,\ldots$,
 \begin{equation}\label{3.8}
 \rho_k({\bf q})=\sum_{{\bf p},\alpha,\alpha'}
 \langle\alpha|\ee^{-\ii kz}|\alpha'\rangle
 a_{\alpha}^\dagger({\bf p})
 a_{\alpha'}^{\vphantom{\dagger}}({\bf p}-{\bf q})
 \end{equation}
 is the mixed Fourier-representation of the local density of electrons,
 \begin{equation}\label{3.9}
 \langle\alpha|\cdots|\alpha'\rangle=
 \int\limits_{-L/2}^{+L/2} \!\!\!\! \dd z \,
 \varphi^*_{\alpha}(z)\cdots\varphi^{\vphantom{*}}_{\alpha'}(z).
 \end{equation}

 It is worth noting that in the equation (\ref{3.10}),
 there are no terms with ${\bf q}=0$,
 due to the electroneutrality condition~(\ref{2.1}).

 The Hamiltonian in the form (\ref{3.10}) is convenient to calculate
 the thermodynamic potential by the functional integration method.

 \section{Thermodynamic potential}

 \subsection{Functional representation}

 The grand partition function,
 \begin{equation}\label{4.1}
 \Xi=\Sp\exp\big[-\beta(H-\mu N)\big],
 \end{equation}
 that determines the thermodynamic potential of the system,
 \begin{equation}\label{OmegaN0}
  \Omega=-\frac1\beta\ln\Xi,
 \end{equation}
 and other thermodynamic functions,
 in the interaction representation becomes,
 \begin{equation}\label{4.2}
 \Xi=\Xi_0\exp\Big(\frac{\beta}{2S}
     \langle N\rangle_0\sum_{\mathbf{q}\neq0}\nu({\bf q},0)
    \Big)\Xi_{\mathrm{int}},
 \end{equation}
 where $\Xi_0=\Sp\exp\big(-\beta(H_0-\mu N)\big)$,
 $H_0=\sum\limits_{{\bf p},\alpha}E_\alpha({\bf p})
 a_\alpha^\dagger({\bf p})a_\alpha^{\vphantom{\dagger}}({\bf p})$
 is the Hamiltonian of non-interacting system,
 $\mu$ is the chemical potential,
 \begin{equation}\label{averPoH0}
 \langle\ldots\rangle_0=\frac1{\Xi_0}\Sp
 \big(\ee^{-\beta (H_0-\mu N)}\ldots\big),
 \end{equation}
 \[
  \langle N\rangle_0=
  \sum\limits_{{\bf p},\alpha}
  \left\langle a_\alpha^\dagger({\bf p})
  a_\alpha^{\vphantom{\dagger}}({\bf p})\right\rangle_0=
  \sum\limits_{{\bf p},\alpha}
  n_\alpha({\bf p}),
 \]
 \[
  n_{\alpha}({\bf p})=\frac1{\ee^{\beta(E_{\alpha}({\bf p})-\mu)}+1}
 \]
 is the Fermi-Dirac distribution,
 \[
  \Xi_{\mathrm{int}}=\langle\mathcal{S}(\beta)\rangle_0,
 \]
 \begin{equation}\label{4.4}
 \mathcal{S}(\beta)=\mathrm{T}\exp\left[-\frac1{2SL}\int\limits_0^\beta \!\! \dd \beta'
 \sum_{\mathbf{q}\neq0}\sum_k\nu_k({\bf q})
  \rho_k({\bf q},\beta')\rho_{-k}(-{\bf q},\beta') \right],
 \end{equation}
 \begin{equation}\label{4.5}
 \rho_k({\bf q},\beta')=
 {\rm e}^{\beta'(H_0-\mu N)}\rho_k({\bf q}){\rm e}^{-\beta'(H_0-\mu N)},
 \end{equation}
 $\mathrm{T}$ is the symbol of chronological ordering of ``times'' $\beta=1/\theta$,
 $\theta$ is the thermodynamic temperature.

 For further calculations it is convenient to switch to spectral representation,
 \begin{equation}\label{4.6}
 \rho_k({\bf q},\nu)=\frac1\beta\int\limits_0^\beta \!
 \dd \beta' \ee^{\ii\nu\beta'}\rho_k({\bf q},\beta'),
 \end{equation}
 \begin{equation}\label{4.7}
 \rho_k({\bf q},\beta')=\sum_\nu{\rm e}^{-\ii\nu\beta'}\rho_k({\bf q},\nu),
 \end{equation}
 where $\nu=\frac{2\pi}{\beta}n$ ($n=0,\pm1,\pm2,\ldots$) are Bose frequencies.
 Then (\ref{4.4}) becomes,
 \begin{equation}\label{4.8}
 \mathcal{S}(\beta)=\mathrm{T}\exp\left[-\frac1{2SL}
 \sum_{\mathbf{q}\neq0}\sum_k\sum_\nu\nu_k({\bf q})
  \rho_k({\bf q},\nu)\rho_{-k}(- \mathbf{q},-\nu) \right].
 \end{equation}

 In order to simplify the approximation $\mathcal{S}(\beta)$ according to (\ref{averPoH0})
 we switch to functional representation for $\mathcal{S}(\beta)$ \cite{Vakar1,Kostr1},
 using Stratonovich-Hubbard identity \cite{Bellman},
 \begin{equation}\label{4.9}
   \exp\left[-\frac12\,\mathbf{y}^\mathrm{T}\mathbb{A}\mathbf{y}\right]\!=
    (\det{\mathbb A})^{-1/2}\!\!\!
   \int\limits_{-\infty}^{+\infty}\!(\dd\mathbf{x})
    \exp\!\left[-\frac12\,\mathbf{x}^\mathrm{T}{\mathbb A}^{-1}\mathbf{x} +
   \ii\, \mathbf{x}^\mathrm{T} \mathbf{y} \right],
 \end{equation}
 where,
 ${(\dd\mathbf{x})=\prod\limits_{k=1}^n \frac{\dd x_k}{\sqrt{2\pi}}}$,
 ${\mathbf{y}^\mathrm{T}=(y_1,\ldots,y_n)}$,
 ${\mathbf{x}^\mathrm{T}=(x_1,\ldots,x_n)}$,
 ${\mathbb A}$ is the positively defined matrix.
 Then, for (\ref{4.8}) we obtain,
 \begin{eqnarray}\label{4.10}
   \mathcal{S}(\beta)&=\prod_{\mathbf{q}\neq0}\prod_k\prod_\nu\big({\textstyle\frac\beta{SL}}\nu_k({\bf q})\big)^{-1/2}\nonumber\\
     &\quad\times\int\!(\dd\omega)\exp\!\!\left[-\frac12\sum_{\mathbf{q}\neq0}\sum_k\sum_\nu
              \big({\textstyle\frac\beta{SL}}\nu_k({\bf q})\big)^{-1}\!
              \omega_k({\bf q},\nu)\omega_{-k}(-\mathbf{q},-\nu)\right]\nonumber\\
     &\quad\times\mathrm{T}\exp\!
              \left[\ii \sum_{\mathbf{q}\neq0}\sum_k\sum_\nu\omega_k({\bf q},\nu) \rho_k({\bf q},\nu)\right],
 \end{eqnarray}
 where $(\dd\omega)$ is the element of the phase space,
 \begin{equation*}
 (\dd\omega)=\prod_{{\bf q}>0}\prod_{k\geqslant 0}\prod_{\nu\geqslant 0}
 \frac{\dd\omega^c_k({\bf q},\nu)}{\sqrt{\pi}}
 \frac{\dd\omega^s_k({\bf q},\nu)}{\sqrt{\pi}},
 \end{equation*}
 \begin{equation*}
 \omega_k({\bf q},\nu)=\omega^c_k({\bf q},\nu)+\ii\omega^s_k({\bf
 q},\nu),
 \end{equation*}
 \begin{equation*}
 \omega^c_k({\bf q},\nu)=\omega^c_{-k}(-{\bf q},-\nu),
 \end{equation*}
 \begin{equation*}
 \omega^s_k({\bf q},\nu)=-\omega^s_{-k}(-{\bf q},-\nu).
 \end{equation*}

 Note, that due to the fact that operator variables $\rho_k({\bf q},\nu)$
 are under sign $\mathrm{T}$-ordering in (\ref{4.10}),
 it is impossible to perform integration by $\beta'$ in Eq. (\ref{4.6}).

 By making average of $\mathcal{S}(\beta) $ according to (\ref{averPoH0}), we obtain,
 \begin{equation}\label{4.15}
  \Xi_{\mathrm{int}}= \big\langle \mathcal{S}(\beta)\big\rangle_0=
   \prod_{\mathbf{q}\neq0}\prod_k\prod_\nu
   \big(\textstyle\frac\beta{S L}\nu_k({\bf q})\big)^{-1/2}
   \!\! \displaystyle\int\!(\dd\omega)J(\omega),
 \end{equation}
 where,
 \begin{eqnarray}\label{4.15a}
  \fl J(\omega)=\exp\!\Bigg[-\frac12\sum_{\mathbf{q}\neq0}\sum_k
 \sum_\nu
 \big(\textstyle\frac\beta{S L}\nu_k({\bf q})\big)^{-1}
 \omega_k({\bf q},\nu)\omega_{-k}(-{\bf q},-\nu)\Bigg]\nonumber\\
  \fl\times
 \exp\Bigg[\sum_{n\geqslant2}\frac1{n!}
 \sum_{
  \begin{array}{c}
    \scriptstyle
    \mathbf{q}_1\neq0,k_1,\nu_1\\ [-3mm]
    \scriptstyle\ldots\ldots\ldots\ldots\ldots\\[-3mm]
    \scriptstyle\mathbf{q}_n\neq0,k_n,\nu_n
  \end{array}}
 {\mathfrak M}^0_{k_1,\ldots,k_n}(\mathbf{q}_1,\nu_1,\ldots,\mathbf{q}_n,\nu_n)
 \omega_{k_1}(\mathbf{q}_1,\nu_1)\ldots\omega_{k_n}(\mathbf{q}_n,\nu_n)\Bigg],
 \end{eqnarray}
\begin{eqnarray*}
  {\mathfrak M}^0_{k_1,\ldots,k_n}(\mathbf{q}_1,\nu_1,\ldots,\mathbf{q}_n,\nu_n)
  &=\ii^n \langle \mathrm{T}\rho_{k_1}(\mathbf{q}_1,\nu_1)\ldots\rho_{k_n}(\mathbf{q}_n,\nu_n)\rangle_{0,\mathrm{c}}\\
  &\sim\kronek_{\mathbf{q}_1+\mathbf{q}_2+\ldots+\mathbf{q}_n,0}\,
       \kronek_{\nu_1+\nu_2+\ldots+\nu_n,0}
\end{eqnarray*}
 are the so-called irreducible mean values (cumulants),
 $\kronek$ is the Kronecker delta,
 ${\mathfrak M}^0_{k}(\mathbf{q},\nu)\equiv0$,
 because ${\bf q}\ne0$.

 In general terms, the calculation of integral (\ref{4.15}) is a complicated problem due to the exponential index
 having terms with $n\geqslant3$.
 Their neglectance results in the Gaussian approximation (or the so-called random phase approximation).
 As a rule, calculation of this integral is done by means of series expansion of the non-Gaussian
 part of the integral (\ref{4.15a}) with subsequent averaging with Gaussian distribution and
 partial summing up of the terms which give the most important contribution.
 In contrast,
 in \cite{CMP2003} it was shown that the integrand $J(\omega)$ can be approximated
 by a Gaussian form $J_\mathrm{G}(\omega)$,
 introducing the unknown function $D_{k_1,k_2}(\mathbf{q},\nu)$:
 \begin{equation}\label{JG}
  \fl J_\mathrm{G}(\omega)
  =\exp\!\Bigg[-\frac12\sum_{\mathbf{q}\neq0}\sum_{k_1,k_2}\sum_\nu
     \big({\textstyle\frac\beta{S}}g_{k_1,k_2}({\bf q},\nu)\big)^{-1}\!
     \omega_{k_1}({\bf q},\nu)\omega_{k_2}(-{\bf q},-\nu)
   \Bigg],
 \end{equation}
 where $g_{k_1,k_2}({\bf q},\nu)$ is the Fourier-transform of effective inter-electron interaction,
 \begin{equation}\label{matrEqForG}
   \big({\textstyle\frac\beta{S}}g_{k_1,k_2}({\bf q},\nu)\big)^{-1}
    =
   \big({\textstyle\frac\beta{S L}}\nu_{k_1}({\bf q})\big)^{-1}\kronek_{k_1+k_2,0}
    -
   D_{k_1,k_2}(\mathbf{q},\nu).
 \end{equation}

 We will seek the unknown function $D_{k_1,k_2}(\mathbf{q},\nu)$ from the condition
 that the mean value of $\omega_{k_1}({\bf q},\nu)\omega_{k_2}(-{\bf q},-\nu)$,
 calculated with the distribution $J(\omega)$,
 is equal to the mean value,
 obtained with Gaussian distribution $J_\mathrm{G}(\omega)$,
 namely from the condition,
 \begin{equation}\label{bb3}
  \overline{\omega_{k_1}({\bf q},\nu)\omega_{k_2}(-{\bf q},-\nu)}=
  \big\langle\omega_{k_1}({\bf q},\nu)\omega_{k_2}(-{\bf q},-\nu)\big\rangle_\mathrm{G},
 \end{equation}
 where we have introduced the following notation,
 \[
  \overline{\vphantom{A}\ldots\vphantom{A}}=
  \frac{\displaystyle\int\!(\dd\omega)J(\omega)\ldots}
        {\displaystyle\int\!(\dd\omega)J(\omega)},
 \]
 \begin{equation}\label{bb4}
  \langle\ldots\rangle_\mathrm{G}=
  \frac{\displaystyle\int\!(\dd\omega)J_\mathrm{G}(\omega)\ldots}
        {\displaystyle\int\!(\dd\omega)J_\mathrm{G}(\omega)}.
 \end{equation}

 In Ref. \cite{CMP2003} it is shown that the solution of Eq. (\ref{bb3}) in the matrix form is,
 \begin{equation}\label{bb11}
  {\mathbb D}=\overline{\mathfrak M}\left({\mathbb I}+{\mathbb V}\,\overline{\mathfrak M}\right)^{-1},
 \end{equation}
 where ${\mathbb I}$ is the identity matrix,
 \[
  {\mathbb V}=\|{\textstyle\frac{\beta}{SL}}\nu_{k_1}(\mathbf{q})\kronek_{k_1+k_2,0}\|,\quad
  \overline{\mathfrak M}=\|\overline{\mathfrak M}_{k_1,k_2}(\mathbf{q},\nu,-\mathbf{q},-\nu)\|,
 \]
 \[
  \overline{\mathfrak M}_{k_1,k_2}(\mathbf{q},\nu,-\mathbf{q},-\nu)
  =
  \ii^2\big\langle \mathrm{T}
  \rho_{k_1}({\bf q},\nu)
  \rho_{k_2}(-{\bf q},-\nu)\big\rangle
 \]
 is the two-particle correlator,
 where the averaging is performed with the Hamiltonian of the system,
 \begin{equation}\label{averPoH}
  \langle\ldots\rangle=\frac1{\Xi}\,\Sp
  \big(\ee^{-\beta (H-\mu N)}\ldots\big).
 \end{equation}

 With this approximation the functional representation for $\Omega_{\mathrm{int}}$ is,
 \begin{equation}\label{omegaIntWithoutLambda}
  \Omega_{\mathrm{int}}=-\frac1\beta\ln\Xi_{\mathrm{int}}=-\frac{1}{\beta}\ln
  \prod\limits_{\mathbf{q}\neq0}
  \prod\limits_{\nu}\prod\limits_{k}
  \Big({\textstyle\frac{\beta}{SL}}\nu_k(\mathbf{q})\Big)^{-1/2}\!\!
  \int\!(\dd \omega)J_\mathrm{G}(\omega).
 \end{equation}
 This Gaussian functional integral can be easily calculated.
 The result is,
 \begin{eqnarray}\label{omegaIntWithoutLambda2}
  \Omega_{\mathrm{int}}&=-\frac{1}{\beta}\ln
  \prod\limits_{\mathbf{q}\neq0}
  \prod\limits_{\nu}\prod\limits_{k}
  \Big({\textstyle\frac{\beta}{SL}}\nu_k(\mathbf{q})\Big)^{-1/2}
  \Big(\det{\textstyle\frac{\beta}{S}}g(\mathbf{q},\nu)\Big)^{1/2}\nonumber\\
  &=-\frac{1}{2\beta}
  \sum\limits_{\mathbf{q}\neq0}
  \sum\limits_{\nu}
  \ln
  \frac{\det g(\mathbf{q},\nu)}
  {\prod\limits_{k}{\textstyle\frac{1}{L}}\nu_k(\mathbf{q})}.
 \end{eqnarray}
 From this expression it follows that
 it is necessary to know the effective inter-electron interaction
 for further calculation of $\Omega_{\mathrm{int}}$.
 Its Fourier-transform satisfies the matrix equation (\ref{matrEqForG}).
 This equation can be written as,
 \begin{equation}\label{matrEqForG1}
  \fl {g}_{k_1,k_2}(\mathbf{q},\nu)= \frac1L\nu_{k_1}({\bf q})\kronek_{k_1+k_2,0}+\frac\beta{S L}\sum_k\nu_{k_1}({\bf q}){D}_{-k_1,k}(\mathbf{q},\nu)
  {g}_{k,k_2}(\mathbf{q},\nu).
 \end{equation}
 Because,
 \begin{eqnarray}\label{gFourier}
  g_{k_1,k_2}(\mathbf{q},\nu)&=\frac1{L^2}
  \int\limits_{-\frac L2}^{+\frac L2}\!\!\dd z_1\!\!
  \int\limits_{-\frac L2}^{+\frac L2}\!\!\dd z_2\;
  \ee^{\ii k_1z_1+\ii k_2z_2}g(\mathbf{q},\nu,z_1,z_2),\nonumber\\ \\
  g(\mathbf{q},\nu,z_1,z_2)&=\sum\limits_{k_1,k_2}
  \ee^{-\ii k_1z_1-\ii k_2z_2}g_{k_1,k_2}(\mathbf{q},\nu)\nonumber
 \end{eqnarray}
 and,
 \begin{eqnarray}\label{DFourier}
  D_{k_1,k_2}(\mathbf{q},\nu)&=\frac1{L^2}
  \int\limits_{-\frac L2}^{+\frac L2}\!\!\dd z_1\!\!
  \int\limits_{-\frac L2}^{+\frac L2}\!\!\dd z_2\;
  \ee^{-\ii k_1z_1-\ii k_2z_2}D(\mathbf{q},\nu,z_1,z_2),\nonumber\\ \\
  D(\mathbf{q},\nu,z_1,z_2)&=\sum\limits_{k_1,k_2}
  \ee^{\ii k_1z_1+\ii k_2z_2}D_{k_1,k_2}(\mathbf{q},\nu),\nonumber
 \end{eqnarray}
 the equation (\ref{matrEqForG1}) in $(\mathbf{q},z)$-representation has the form,
 \begin{eqnarray}\label{intEq0}
   \fl g(\mathbf{q},\nu,z_1,z_2)=\nu({\bf q},z_1-z_2)   \nonumber\\
   +\frac\beta{S L^2}\!
                   \int\limits_{-\frac L2}^{+\frac L2} \!\! \dd z \!\!
                   \int\limits_{-\frac L2}^{+\frac L2} \!\! \dd z' \,
                   \nu({\bf q},z_1-z')D(\mathbf{q},\nu,z',z)g(\mathbf{q},\nu,z,z_2),
 \end{eqnarray}
 This integral equation was solved with different approximations in Refs. \cite{CMP2003,JPS2003_1,CMP2006,UFZ2007,FMMIT2013}.

 Consequently,
 the following expression for the thermodynamic potential $\Omega$ (\ref{OmegaN0})
 can be obtained,
 \begin{equation}
 \label{omega}
 \Omega=\Omega_0-\frac{1}{2S}\langle N\rangle_0\sum_{\mathbf{q}\neq0}
 \nu({\bf q},0)
 +\Omega_{\mathrm{int}},
 \end{equation}
 where,
 \begin{equation}\label{omega0}
 \Omega_0=-\frac1\beta\ln\Xi_0
 =-\frac1\beta\sum\limits_{\mathbf{p},\alpha}
 \ln\left[1+\ee^{\beta(\mu-E_\alpha(\mathbf{p}))}\right]
 \end{equation}
 is the thermodynamical potential of non-interacting
 system\footnote{Because the form of the thermodynamic potential $\Omega_0$
                 as a function of $\mu$ coincides with the thermodynamic potential
                 of an ideal electron gas,
                 the thermodynamic potential $\Omega_0$~(\ref{omega0})
                 is called by us as `thermodynamic potential of non-interacting system'
                 though it indirectly takes into account the Coulomb interaction
                 between electrons via the chemical potential~$\mu$ of interacting electrons.
                 The same applies to the internal energy of non-interacting system~$U_0$.},
 \begin{equation}\label{omegaInt}
  \Omega_{\mathrm{int}}=\frac{1}{2\beta}
  \sum\limits_{\mathbf{q}\neq0}
  \sum\limits_{\nu}
  \ln
  \frac{\det g(\mathbf{q},\nu)}
  {\prod\limits_{k}{\textstyle\frac{1}{L}}\nu_k(\mathbf{q})}.
 \end{equation}

 Therefore,
 the calculation of the thermodynamic potential in this approach requires knowledge of
 the effective inter-electron interaction with taking into account the presence of the dividing plane.

 \subsection{The calculation of $\Omega_0$}\label{OmegaNonInteracting}

 Let us now calculate the thermodynamic potential of the non-interacting system,
 \begin{equation}\label{omega1}
 \Omega_0=-\frac1\beta\sum\limits_{\mathbf{p},\alpha}
 \ln\left[1+\ee^{\beta(\mu-E_\alpha(\mathbf{p}))}\right].
 \end{equation}
 Because here $\mu$ is the chemical potential of interacting electrons,
 this expression takes into account the Coulomb interaction indirectly.

 In order to perform the summation by $\mathbf{p}$ and $\alpha$,
 we use the density of states calculated in the Appendix~\ref{DOSapp}.
 Then, the thermodynamic potential is,
 \begin{eqnarray*}
   \fl \Omega_0=-\frac1\beta
       \int\limits_0^\infty\!\dd E\,\rho(E)
       \ln\left[1+\ee^{\beta(\mu-E)}\right]
       =-\frac1\beta
          \frac{SL}2\frac{\sqrt{2}m^{3/2}}{\pi^2\hbar^3}
          \int\limits_0^\infty\!\dd E\,\sqrt{E}\,
          \ln\left[1+\ee^{\beta(\mu-E)}\right]\\
        -\frac1\beta
          S\frac{\sqrt{2}m^{3/2}d}{\pi^2\hbar^3}
          \int\limits_0^\infty\!\dd E\,\sqrt{E}\,
          \ln\left[1+\ee^{\beta(\mu-E)}\right]
          +\frac1\beta
          S\frac{m}{4\pi\hbar^2}
          \int\limits_0^\infty\!\dd E\,
          \ln\left[1+\ee^{\beta(\mu-E)}\right].
 \end{eqnarray*}
 Integrating by parts each of the terms, we get,
 \begin{eqnarray*}
  \fl \Omega_0
       =-\frac{SL}2\frac{2\sqrt{2}m^{3/2}}{3\pi^2\hbar^3}
          \int\limits_0^\infty\!\dd E\,{E}^{3/2}\,
          \frac1{\ee^{\beta(E-\mu)}+1}\\
        -S\frac{2\sqrt{2}m^{3/2}d}{3\pi^2\hbar^3}
          \int\limits_0^\infty\!\dd E\,{E}^{3/2}\,
          \frac1{\ee^{\beta(E-\mu)}+1}
          +S\frac{m}{4\pi\hbar^2}
          \int\limits_0^\infty\!\dd E\,E\,
          \frac1{\ee^{\beta(E-\mu)}+1}.
 \end{eqnarray*}
 In the limit of low temperatures (${\beta\to\infty}$),
 we get the following expression,
 \begin{equation}\label{Omega0Suma}
  \Omega_0=\Omega_{0,\bulk}+\Omega_{0,\surf},
 \end{equation}
 where,
 \begin{equation}\label{omega2bulk}
   \Omega_{0,\bulk}=-\frac{SL}2\frac{4\sqrt{2}m^{3/2}}{15\pi^2\hbar^3}\mu^{5/2}
                   =-\frac{SL}2\frac{\hbar^2}{15m\pi^2}p_\mathrm{F}^5
 \end{equation}
 is the extensive contribution to the thermodynamic potential of the non-interacting system
 (it is proportional to the volume ${SL}$),
 which depends on the Fermi momentum $p_\mathrm{F}$ of interacting electrons
 and,
 \begin{equation}\label{omega2IBMsurf}
   \Omega_{0,\surf}=-S\left(\frac{4\sqrt{2}m^{3/2}d}{15\pi^2\hbar^3}\mu^{5/2}
                     -\frac{m}{8\pi\hbar^2}\mu^2\right)
         =  S\,\frac{\hbar^2p_\mathrm{F}^4}{m\pi^2}
           \left(\frac{\pi}{32}-\frac{d\,p_\mathrm{F}}{15}\right)
 \end{equation}
 is the surface contribution (it is proportional to the area of the dividing plane $S$).

 It should be noted that
 the thermodynamic potential of the non-interacting homogeneous system is
 \begin{equation}\label{omega0unif}
  \Omega_{0,\unif}=-V\,\frac{4\sqrt{2}m^{3/2}}{15\pi^2\hbar^3}\mu^{5/2}=
                  -V\,\frac{\hbar^2}{15m\pi^2}p_\mathrm{F}^5,
 \end{equation}
 which coincides with (\ref{omega2bulk}).

\subsection{The calculation of $\Omega_{\mathrm{int}}$}

 As shown in Eq. (\ref{omegaInt}),
 the calculation of the thermodynamic potential requires to evaluate
 the determinant of the matrix effective inter-electron interaction.
 It is challenging,
 since the order of the matrix is infinite.
 We can use the well-known identity (see, for example, \cite{Kleinert})
 \[
  \ln\det\mathbb{A}=\Sp\ln\mathbb{A},
 \]
 with further expansions in series $\ln\mathbb{A}$,
 next calculating the trace of each matrix term in the series and following summation of the series.

 To avoid this, we propose a different procedure (it is equivalent to the approach outlined above).
 The essence is to build a differential equation for the unknown quantity $\Omega_{\mathrm{int}}$.
 To this end,
 we introduce the function $J_\mathrm {G}(\omega,\lambda)$
 instead of $J_\mathrm{G}(\omega)$ (\ref{JG}),
 which depends on the parameter $\lambda$:
 \begin{equation}\label{JGlambda}
 \fl J_\mathrm{G}(\omega,\lambda)=\exp\!\Bigg[-\frac12\sum_{\mathbf{q}\neq0}\sum_{k_1,k_2}\sum_\nu
     \big({\textstyle\frac\beta{S}}g_{k_1,k_2}({\bf q},\nu,\lambda)\big)^{-1}\!
     \omega_{k_1}({\bf q},\nu)\omega_{k_2}(-{\bf q},-\nu)
   \Bigg],
 \end{equation}
 that depends on the parameter $\lambda$,
 \begin{equation}\label{matrEqForGlambda}
   \big({\textstyle\frac\beta{S}}g_{k_1,k_2}({\bf q},\nu,\lambda)\big)^{-1}
    =
   \big({\textstyle\frac\beta{S L}}\nu_{k_1}({\bf q})\big)^{-1}\kronek_{k_1+k_2,0}
    -
   \lambda\,D_{k_1,k_2}(\mathbf{q},\nu),
 \end{equation}
 moreover $g_{k_1,k_2}({\bf q},\nu)\equiv g_{k_1,k_2}({\bf q},\nu,1)$,
 $J_\mathrm{G}(\omega)\equiv J_\mathrm{G}(\omega,1)$.

 Then, $\Omega_{\mathrm{int}}$ and $\Xi_{\mathrm{int}}$ will depend on this parameter as well,
 \begin{eqnarray}\label{omegaIntLambda}
  \Omega_{\mathrm{int}}(\lambda)&=-\frac{1}{\beta}\ln\Xi_{\mathrm{int}}(\lambda)\nonumber\\
  &=-\frac{1}{\beta}\ln
  \prod\limits_{\mathbf{q}\neq0}
  \prod\limits_{\nu}\prod\limits_{k}
  \big({\textstyle\frac{\beta}{SL}}\nu_k(\mathbf{q})\big)^{-1/2}
  \int\!(\dd \omega)J_\mathrm{G}(\omega,\lambda),
 \end{eqnarray}
 moreover
 \[
  \Omega_{\mathrm{int}}=\Omega_{\mathrm{int}}(1).
 \]

 We need to perform differentiation of $\Omega_{\mathrm{int}}(\lambda)$ with respect to the parameter $\lambda$,
 and arrive at the result,
 \begin{eqnarray*}
   \fl \frac{\dd\Omega_{\mathrm{int}}(\lambda)}{\dd\lambda}
    =-\frac1\beta\frac{\dd}{\dd\lambda}\ln\Xi_{\mathrm{int}}(\lambda)
     =-\frac1\beta\frac1{\Xi_{\mathrm{int}}(\lambda)}\frac{\dd\Xi_{\mathrm{int}}(\lambda)}{\dd\lambda}=\nonumber\\
    =-\frac1\beta\frac1{\Xi_{\mathrm{int}}(\lambda)}
      \prod\limits_{\mathbf{q}\neq0}\prod\limits_{\nu}\prod\limits_{k}
      \big({\textstyle\frac{\beta}{SL}}\nu_k(\mathbf{q})\big)^{-1/2}\times\\
    \quad\times  \frac12\int\!(\dd \omega)J_\mathrm{G}(\omega,\lambda)
   \sum\limits_{
                \begin{array}{c}
                 \scriptstyle \mathbf{q}\neq0,\nu \\[-2mm]
                 \scriptstyle k_1,k_2
                \end{array}
               } \!\!
     D_{k_1,k_2}(\mathbf{q},\nu)
     \omega_{k_1}(\mathbf{q},\nu)\omega_{k_2}(-\mathbf{q},-\nu)=\\
   =-\frac1{2\beta}
    \sum\limits_{
                \begin{array}{c}
                 \scriptstyle \mathbf{q}\neq0,\nu \\[-2mm]
                 \scriptstyle k_1,k_2
                \end{array}
                } \!\!
     D_{k_1,k_2}(\mathbf{q},\nu)
     \big\langle\omega_{k_1}(\mathbf{q},\nu)\omega_{k_2}(-\mathbf{q},-\nu)\big\rangle_\mathrm{G}(\lambda),
 \end{eqnarray*}
 where we have introduced the average,
 \begin{equation}\label{averLambda}
  \langle\ldots\rangle_\mathrm{G}(\lambda)
  =\frac{\displaystyle\int(\dd\omega)J_\mathrm{G}(\omega,\lambda)\ldots}
                           {\displaystyle\int(\dd\omega)J_\mathrm{G}(\omega,\lambda)}.
 \end{equation}
 Both averaging (\ref{bb4}) and (\ref{averLambda}) coincide when in the latter the parameter $\lambda$ is $1$,
 \[
    \langle\ldots\rangle_\mathrm{G}=\langle\ldots\rangle_\mathrm{G}(1)
 \]

 Therefore, $\Omega_{\mathrm{int}}(\lambda)$ satisfies the differential equation of the first order,
 \begin{equation}\label{eqForOmegaInt}
    \frac{\dd\Omega_{\mathrm{int}}(\lambda)}{\dd\lambda}
    =-\frac1{2\beta}\sum\limits_{                \begin{array}{c}
                 \scriptstyle \mathbf{q}\neq0,\nu \\[-2mm]
                 \scriptstyle k_1,k_2
                \end{array}} \!\!
     D_{k_1,k_2}(\mathbf{q},\nu)
     \big\langle\omega_{k_1}(\mathbf{q},\nu)\omega_{k_2}(-\mathbf{q},-\nu)\big\rangle_\mathrm{G}(\lambda).
 \end{equation}
 In order to obtain an unambiguous solution of the differential equation of the first order
 it must be supplemented by the single additional condition.
 It is worth noting that,
 \begin{eqnarray}\label{pochUmovaForOmegaInt}
   \Omega_{\mathrm{int}}(0)&=-\frac{1}{\beta}\ln
  \prod\limits_{\mathbf{q}\neq0}
  \prod\limits_{\nu}\prod\limits_{k}
  \big({\textstyle\frac{\beta}{SL}}\nu_k(\mathbf{q})\big)^{-1/2}\!\!\!\nonumber\\
  &\quad\times\int\!(\dd \omega)
   \exp\!\bigg[\!
    -\frac12\sum\limits_{\mathbf{q}\neq0,\nu,k}\!\!
     \!\!\big({\textstyle\frac{\beta}{SL}}\nu_k(\mathbf{q})\big)^{-1}\!
     \omega_k(\mathbf{q},\nu)
     \omega_{-k}(-\mathbf{q},-\nu)
    \bigg]\nonumber\\
   &= \frac{1}{\beta}\ln1=0.
 \end{eqnarray}

 We easily find the solution of the Cauchy problem (\ref{eqForOmegaInt}), (\ref{pochUmovaForOmegaInt}):
 \begin{equation}\label{solutionForOmegaInt}
    \fl \Omega_{\mathrm{int}}\equiv \Omega_{\mathrm{int}}(1)=-\frac1{2\beta}\sum\limits_{                \begin{array}{c}
                 \scriptstyle \mathbf{q}\neq0,\nu \\[-2mm]
                 \scriptstyle k_1,k_2
                \end{array}} \!\!
     D_{k_1,k_2}(\mathbf{q},\nu)
     \int\limits_0^1\!
     \big\langle\omega_{k_1}(\mathbf{q},\nu)\omega_{k_2}(-\mathbf{q},-\nu)\big\rangle_\mathrm{G}(\lambda)
     \dd\lambda.
 \end{equation}
 Average of $\omega_{k_1}(\mathbf{q},\nu)\omega_{k_2}(-\mathbf{q},-\nu)$
 with the Gaussian distribution $J_\mathrm{G}(\omega,\lambda)$
 yields,
 \begin{equation}\label{averOmegaOmegaLambda}
  \big\langle\omega_{k_1}(\mathbf{q},\nu)\omega_{k_2}(-\mathbf{q},-\nu)\big\rangle_\mathrm{G}(\lambda)
  =\frac\beta S g_{k_1,k_2}(\mathbf{q},\nu,\lambda).
 \end{equation}

 Thus, we have obtained a convenient expression for the calculation of $\Omega_{\mathrm{int}}$:
 \begin{equation}\label{solutionForOmegaInt2}
    \Omega_{\mathrm{int}}=
     -\frac1{2S}\sum\limits_{                \begin{array}{c}
                 \scriptstyle \mathbf{q}\neq0,\nu \\[-2mm]
                 \scriptstyle k_1,k_2
                \end{array}} \!\!
     D_{k_1,k_2}(\mathbf{q},\nu)
     \int\limits_0^1\!
     g_{k_1,k_2}(\mathbf{q},\nu,\lambda)
     \dd\lambda.
 \end{equation}
 Using the relations (\ref{gFourier})
 (which holds for the effective inter-electron interaction dependent on $\lambda$)
 and (\ref{DFourier}),
 we get,
 \begin{equation}\label{solutionForOmegaInt3}
    \Omega_{\mathrm{int}}=
     -\frac1{2SL^2}\sum\limits_{\mathbf{q}\neq0,\nu}
     \int\limits_{-\frac L2}^{+\frac L2}\!\!\dd z_1\!\!
     \int\limits_{-\frac L2}^{+\frac L2}\!\!\dd z_2\;
     D(\mathbf{q},\nu,z_1,z_2)
     \int\limits_0^1\!
     g(\mathbf{q},\nu,z_1,z_2,\lambda)
     \dd\lambda.
 \end{equation}

 Further evaluation of $\Omega_{\mathrm{int}}$ using this formula
 should be carried out numerically.
 In order to obtain the analytical results, we make the following approximation:
 \begin{itemize}
   \item $D\approx{\mathfrak M}^0$, namely we apply the random phase approximation;
   \item $g(\mathbf{q},\nu,z_1,z_2,\lambda)\approx g(\mathbf{q},0,z_1,z_2,\lambda)\equiv
          g(\mathbf{q},z_1,z_2,\lambda)$,
         i.e. we neglect the dependence of the effective inter-electron interaction on Bose frequency $\nu$.
 \end{itemize}
 Then, the expression for $\Omega_{\mathrm{int}}$ is simplified,
 \begin{equation}\label{solutionForOmegaInt4}
   \fl \Omega_{\mathrm{int}}\approx
     -\frac1{2SL^2}\sum\limits_{\mathbf{q}\neq0}
     \int\limits_{-\frac L2}^{+\frac L2}\!\!\dd z_1\!\!
     \int\limits_{-\frac L2}^{+\frac L2}\!\!\dd z_2
     \sum\limits_{\nu}{\mathfrak M}^0(\mathbf{q},\nu,z_1,z_2)
     \int\limits_0^1\!
     g(\mathbf{q},z_1,z_2,\lambda)
     \dd\lambda.
 \end{equation}
 In this expression the summation by the frequency $\nu$ only applies to the function
  \begin{equation}\label{corrFunc}
   \fl{\mathfrak M}^0(\mathbf{q},\nu,z_1,z_2)=\frac{L^2}{\beta}\!\!\sum_{{\bf p},\alpha_1,\alpha_2}\!
       \frac{n_{\alpha_1}({\bf p})-n_{\alpha_2}({\bf p}-\mathbf{q})}
            {-\ii\nu+E_{\alpha_1}({\bf p})-E_{\alpha_2}({\bf p}-\mathbf{q})}
      \varphi^*_{\alpha_1}\!(z_1) \varphi^{\vphantom{*}}_{\alpha_2}\!(z_1)
      \varphi^*_{\alpha_2}\!(z_2) \varphi^{\vphantom{*}}_{\alpha_1}\!(z_2)  \nonumber
 \end{equation}
 and can be performed analytically,
 \begin{eqnarray}\label{sumPoNuM0}
 \fl \sum\limits_{\nu}{\mathfrak M}^0(\mathbf{q},\nu,z_1,z_2)=-L^2
  \sum_{{\bf p},\alpha_1}n_{\alpha_1}({\bf p})
  |\varphi_{\alpha_1}\!(z_1)|^2 \dirac(z_1-z_2)\nonumber\\
  +L^2 \sum_{{\bf p},\alpha_1,\alpha_2}n_{\alpha_1}({\bf p})n_{\alpha_2}({\bf p}-\mathbf{q})
  \varphi^*_{\alpha_1}\!(z_1) \varphi^{\vphantom{*}}_{\alpha_2}\!(z_1)
      \varphi^*_{\alpha_2}\!(z_2) \varphi^{\vphantom{*}}_{\alpha_1}\!(z_2).
 \end{eqnarray}
 The effective inter-electron interaction $g(\mathbf{q},z_1,z_2,\lambda)$
 is the solution of the integral equation:
 \begin{eqnarray}\label{intEq1}
    g(\mathbf{q},z_1,z_2,\lambda)&=\nu({\bf q},z_1-z_2)\\
     &\quad           +\frac\beta{S L^2}\,\lambda\!\!\!\!
                   \int\limits_{-L/2}^{+L/2} \!\!\!\! \dd z \!\!
                   \int\limits_{-L/2}^{+L/2} \!\!\!\! \dd z' \,
                   \nu({\bf q}|z_1-z'){\mathfrak M}^0(\mathbf{q},0,z',z)g(\mathbf{q},z,z_2,\lambda).\nonumber
 \end{eqnarray}

 Substituting Eq. (\ref{sumPoNuM0}) into Eq. (\ref{solutionForOmegaInt4}),
 and Eq. (\ref{solutionForOmegaInt4}) in Eq. (\ref{omega}),
 we find the thermodynamic potential,
 \begin{eqnarray}\label{omega5}
    \fl\Omega=\Omega_0-\frac1{2S}\langle N\rangle_0\sum\limits_{\mathbf{q}\neq0}\nu(\mathbf{q},0)
    +\frac1{2S}\sum\limits_{\mathbf{q}\neq0}\sum\limits_{\mathbf{p},\alpha}n_\alpha(\mathbf{p})
      \int\limits_{-\frac L2}^{+\frac L2}\!\!\dd z\,|\varphi_\alpha(z)|^2
      \int\limits_0^1\!g(\mathbf{q},z,z,\lambda)\dd\lambda\\
  \fl  -\frac1{2S}\sum\limits_{\mathbf{q}\neq0}\sum\limits_{\mathbf{p},\alpha_1,\alpha_2}\!\!
      n_{\alpha_1}(\mathbf{p}) n_{\alpha_2}(\mathbf{p}-\mathbf{q})
      \int\limits_{-\frac L2}^{+\frac L2}\!\!\dd z_1\!\!
      \int\limits_{-\frac L2}^{+\frac L2}\!\!\dd z_2\,
      \varphi^*_{\alpha_1}\!(z_1) \varphi^{\vphantom{*}}_{\alpha_2}\!(z_1)
      \varphi^*_{\alpha_2}\!(z_2) \varphi^{\vphantom{*}}_{\alpha_1}\!(z_2)
      \int\limits_0^1\!g(\mathbf{q},z_1,z_2,\lambda)\dd\lambda.\nonumber
 \end{eqnarray}

 Taking into account the expressions for one- and two-particle
 distribution functions of electrons in the semi-infinite jellium \cite{JPS2003_2},
 \begin{equation}\label{unarna}
    F_1^0(z)=\frac{V}{S\langle N\rangle_0}\sum\limits_{\mathbf{p},\alpha}|\varphi_\alpha(z)|^2n_\alpha(\mathbf{p}),
 \end{equation}
 \begin{eqnarray}\label{binarna}
    \fl F_2^0(\mathbf{r}_{||},z_1,z_2)=F_1^0(z_1)F_1^0(z_2)\nonumber\\
    -    \frac{V^2}{S^2\langle N\rangle_0^2}\!
    \sum\limits_{                \begin{array}{c}
                 \scriptstyle \mathbf{p},\alpha_1 \\[-2mm]
                 \scriptstyle \mathbf{p}',\alpha_2
                \end{array}}\!\!
    \ee^{\ii\mathbf{p}\mathbf{r}_{||}}
    n_{\alpha_1}(\mathbf{p}') n_{\alpha_2}(\mathbf{p}'-\mathbf{p})
    \varphi^*_{\alpha_1}\!(z_1) \varphi^{\vphantom{*}}_{\alpha_2}\!(z_1)
      \varphi^*_{\alpha_2}\!(z_2) \varphi^{\vphantom{*}}_{\alpha_1}\!(z_2),
 \end{eqnarray}
 the thermodynamic potential can be represented as,
 \begin{eqnarray}\label{omega6}
    \fl\Omega=\Omega_0-\frac1{2S} \langle N\rangle_0\sum\limits_{\mathbf{q}\neq0}\nu(\mathbf{q},0)
    +\frac1{2}\frac{\langle N\rangle_0S}{V}\int\limits_{-\frac L2}^{+\frac L2}\!\!\dd z\,F_1^0(z)
      \int\limits_0^1\!\dd\lambda \,g(\mathbf{r}_{||},z,z,\lambda)\big|_{\mathbf{r}_{||}=0}\nonumber\\
   \fl +\frac1{2}\frac{\langle N\rangle_0^2S}{V^2}
      \int\limits_{S}\!\dd \mathbf{r}_{||}\!\!
      \int\limits_{-\frac L2}^{+\frac L2}\!\!\dd z_1\!\!
      \int\limits_{-\frac L2}^{+\frac L2}\!\!\dd z_2\,
      \left(
        F_2^0(\mathbf{r}_{||},z_1,z_2)-F_1^0(z_1)F_1^0(z_2)
      \right) \int\limits_0^1\!\dd\lambda \,g(\mathbf{r}_{||},z_1,z_2,\lambda),
 \end{eqnarray}
 where,
 \begin{equation}\label{gKoord}
  g(\mathbf{r}_{||},z_1,z_2,\lambda)=\frac1S
  \sum\limits_{\mathbf{q}}\ee^{\ii\mathbf{q}\mathbf{r}_{||}}g(\mathbf{q},z_1,z_2,\lambda)
 \end{equation}
 is the effective inter-electron interaction in the coordinate representation,
 which depends on the parameter $\lambda$.

 It should be noted that
 the expressions (\ref{unarna}) and (\ref{binarna}) coincide
 by form with the expressions for the distribution functions of electrons
 without the Coulomb interaction, but these distribution functions
 depend on the chemical potential~$\mu$ of interacting electrons.

 \subsubsection*{Thermodynamic potential in the case of the infinite barrier model.}

 The form of the surface potential $V_\surf(z)$ must be specified
 to perform further calculation of the thermodynamic potential according
 to the expressions (\ref{omega5}) or (\ref{omega6}).
 We use the infinite barrier model for the surface potential,
 namely,
 \begin{equation}\label{poverhPot}
  V_\surf(z)=\left\{
   \begin{array}{cl}
     \infty, & z>d, \\
     0, & z<d.
   \end{array}
  \right.
 \end{equation}
 The wave functions and the corresponding energy levels for the model are,
 \begin{equation}\label{waveFunctionIBM}
    \varphi_\alpha(z)=\frac{2}{\sqrt{L+2d}}
    \left\{
     \begin{array}{ll}
       \sin\big(\alpha(d-z)\big), & z\leqslant d, \\[2mm]
       0, & z>d,
     \end{array}
    \right.\quad
  \varepsilon_\alpha=\frac{\hbar^2\alpha^2}{2m},
 \end{equation}
 where,
 \begin{equation}\label{eqForAlfa}
  \alpha=\frac{\pi n}{\left(\frac{L}{2}+d\right)},\;
  n=1,2,3,\ldots.
 \end{equation}

 The one-particle distribution function of the model is,
 \begin{equation}\label{unarnaIBM}
  F_1^0(z)=
  \left[
   1+\frac{3\cos\big(2p_{\mathrm{F}}(d-z)\big)}{\big(2p_{\mathrm{F}}(d-z)\big)^2}
   -\frac{3\sin\big(2p_{\mathrm{F}}(d-z)\big)}{\big(2p_{\mathrm{F}}(d-z)\big)^3}
  \right]
  \heav(d-z).
 \end{equation}
 This expression for $d=0$ coincides by form with the one-particle distribution function
 without Coulomb interaction \cite{Stratton,Sugiyama2,Newns},
 but the expression (\ref{unarnaIBM}) takes into account the Coulomb interaction
 through the Fermi momentum $p_\mathrm{F}$ of interacting electrons.

 Using the technique to solve the integral equation (\ref{intEq1})
 (see \cite{JPS2003_1,CMP2006}), 
 we obtain the following expression for the effective inter-electron interaction $g(\mathbf{q},z_1,z_2,\lambda)$,
 \begin{eqnarray*}
  g({q},z_1\leqslant d,z_2\leqslant d,\lambda)&=\frac{2\pi e^2}{Q(\lambda)}\!
                      \left[
                        \ee^{-Q(\lambda)|z_1-z_2|}+\!\frac{Q(\lambda)-q}{Q(\lambda)+q}\ee^{Q(\lambda)(z_1+z_2-2d)}
                      \right]\!\!,\\
  g({q},z_1\geqslant d,z_2\geqslant d,\lambda)&=\frac{2\pi e^2}{q}
                      \left[
                        \ee^{-q|z_1-z_2|}-\frac{Q(\lambda)-q}{Q(\lambda)+q}\ee^{-q(z_1+z_2-2d)}
                      \right]\!\!,\\
  g({q},z_1\geqslant d,z_2\leqslant d,\lambda)&=\frac{4\pi e^2}{Q(\lambda)+q}
                        \,\ee^{Q(\lambda)(z_2-d)-q(z_1-d)},\\
  g({q},z_1\leqslant d,z_2\geqslant d,\lambda)&=\frac{4\pi e^2}{Q(\lambda)+q}
                        \,\ee^{Q(\lambda)(z_1-d)-q(z_2-d)},
 \end{eqnarray*}
 where,
 \[
  Q(\lambda)=\sqrt{q^2+\lambda\,\varkappa_{\mathrm{TF}}^2\lindhard\big(\textstyle\frac{q}{2p_{\mathrm{F}}}\big)},
 \]
 \[
  \lindhard(x)=\frac12+\frac{1-x^2}{4x}\ln\left|\frac{1+x}{1-x}\right|,
 \]
 ${\varkappa_{\mathrm{TF}}=\sqrt{\frac{4}{\pi}\frac{p_\mathrm{F}}{a_\mathrm{B}}}}$
 is the inverse Thomas-Fermi radius of screening, and $a_\mathrm{B}$ is the Bohr radius.

 After the summation by the momenta $\mathbf{p}$ in (\ref{omega5}),
 we get,
 \begin{equation}\label{omega7}
    \Omega=\Omega_\bulk+\Omega_\surf,
 \end{equation}
 where the first term is the extensive contribution to the thermodynamic potential
 (it is proportional to the volume of the system ${SL}$),
 the second term is the surface contribution (it is proportional to the area of the dividing plane $S$).
 The extensive contribution to the thermodynamic potential is,
 \begin{equation}\label{omegaBulk1}
    \Omega_\bulk=\Omega_{0,\bulk}+\Delta\Omega_\bulk,
 \end{equation}
 where,
 \begin{equation}\label{omegaBulk1a}
   \frac{\Omega_{0,\bulk}}{{SL}/2}=-\frac{\hbar^2}{15m\pi^2}p_\mathrm{F}^5,
 \end{equation}
 is the extensive contribution to the thermodynamic potential of the non-interacting system
 per unit volume (see the expression (\ref{omega2bulk})).
 This contribution depends on the Fermi momentum $p_{\mathrm{F}}$ of interacting electrons.
 $\Delta\Omega_\bulk$ has the form,
 \begin{eqnarray}\label{omegaBulk1b}
    \frac{\Delta\Omega_{\bulk}}{{SL}/2}&=\frac{e^2p_\mathrm{F}^3}{6\pi^2}
      \int\limits_0^\infty\!\dd q
      \Bigg[
       \int\limits_0^1\!\dd \lambda\,\frac{q}{Q(\lambda)}-1
      \Bigg]\nonumber\\
      &\quad-\frac{e^2}{2\pi^4}
      \int\limits_0^\infty\!\!\dd q\,q\!
      \int\limits_0^\infty\!\!\dd \alpha_1\!\!
      \int\limits_0^\infty\!\!\dd \alpha_2
      \widetilde{J}(q,\alpha_1,\alpha_2)\!
      \int\limits_0^1\!\dd \lambda\,\,
      g_1(q,\alpha_1,\alpha_2,\lambda),
 \end{eqnarray}
 where,
  \begin{equation*}
  \widetilde{J}(q,\alpha_1,\alpha_2)=
    \left\{
      \begin{array}{cl}
        \left\{
         \begin{array}{ll}
            \pi c_1^2, & c_2>c_1, \\
            \pi c_2^2, & c_1>c_2,
         \end{array}
        \right. & 0\leqslant q<|c_1-c_2|,\\[4mm]
        f(c_1,c_2,q) + f(c_2,c_1,q),& |c_1-c_2|\leqslant q< c_1+c_2, \\[4mm]
        0,& q\geqslant c_1+c_2,
      \end{array}
    \right.
 \end{equation*}
 \[
  c_1=\sqrt{p_\mathrm{F}^2-\alpha_1^2},\quad
  c_2=\sqrt{p_\mathrm{F}^2-\alpha_2^2},
 \]
 \begin{equation*}
   \fl f(c_1,c_2,q)=c_1^2\left(\frac{\pi}{2}-\arcsin\frac{c_1^2-c_2^2+q^2}{2qc_1}\right)
   -\frac{c_1^2-c_2^2+q^2}{2q}\sqrt{c_1^2-\frac{(c_1^2-c_2^2+q^2)^2}{4q^2}},
 \end{equation*}
 \begin{equation*}
    g_1(q,\alpha_1,\alpha_2,\lambda)=\frac{Q^2(\lambda)+\alpha_1^2+\alpha_2^2}
             {\big(Q^2(\lambda)+\alpha_1^2+\alpha_2^2\big)^2-4\alpha_1^2\alpha_2^2}.
 \end{equation*}

 It should be noted that
 the dividing plane has no effect on the expression given by Eq. (\ref{omegaBulk1})
 and actually this expression is the thermodynamic potential of the homogeneous system per unit volume.
 However, the thermodynamic potential of the homogeneous system can be calculated
 in a similar manner,
 based on the Hamiltonian of the infinite jellium (\ref{HamiltonianUnif}).
 These calculations are much simpler and at the similar level of approximations we obtain,
 \begin{eqnarray}\label{OmegaUnif2}
    \frac{\Omega_\unif}{V}&=-\frac{\hbar^2}{15m\pi^2}p_\mathrm{F}^5
     +\frac{p_\mathrm{F}^3}{12\pi^4}
      \int\limits_0^\infty\!\dd q\,q^2
      \Bigg[
       \int\limits_0^1\!\dd \lambda\,g_\unif(q,\lambda)-\nu(q)
      \Bigg]\nonumber\\
     &\quad-\frac{p_\mathrm{F}^3}{12\pi^4}
      \int\limits_0^{\infty}\!\dd q\,q^2\!
      \widetilde{J}_\unif(q)
      \int\limits_0^1\!\dd \lambda\,\,
      g_\unif(q,\lambda),
 \end{eqnarray}
 where,
 \begin{equation*}
    \widetilde{J}_\unif(q)
    =\left\{
     \begin{array}{cc}
       1-\frac34\frac{q}{p_\mathrm{F}}+\frac1{16}\frac{q^3}{p_\mathrm{F}^3}, & q<2p_\mathrm{F}, \\[3mm]
       0, & q\geqslant2p_\mathrm{F},
     \end{array}
      \right.
 \end{equation*}
 $\nu(q)=\frac{4\pi e^2}{q^2}$ is the three-dimensional Fourier-transform of the Coulomb interaction,
 \[
  g_\unif(q,\lambda)=\frac{4\pi e^2}{q^2+\lambda\,\varkappa^2_{\mathrm{TF}}L\big(\textstyle\frac{q}{2p_\mathrm{F}}\big)}
 \]
 is the three-dimensional Fourier-transform of the effective inter-electron interaction of the homogeneous system,
 that depends on the parameter $\lambda$.

 Integration by the parameter $\lambda$ in the expression given by Eq. (\ref{OmegaUnif2}) can be easily performed
 and as a result we obtain,
 \begin{equation}\label{OmegaUnif3}
   \fl \frac{\Omega_\unif}{V}=-\frac{\hbar^2}{15m\pi^2}p_\mathrm{F}^5-
    \frac{e^2p_\mathrm{F}^3}{3\pi^3}
      \int\limits_0^\infty\!\dd q\,
      \Bigg[1-
       \frac{q^2\big(1-\widetilde{J}_\unif(q)\big)}
       {\varkappa^2_{\mathrm{TF}}L\big(\textstyle\frac{q}{2p_\mathrm{F}}\big)}
       \ln\left(1+\frac{\varkappa^2_{\mathrm{TF}}}{q^2}L
       \big(\textstyle\frac{q}{2p_\mathrm{F}}\big)\right)
      \Bigg].
 \end{equation}
 Although the expressions (\ref{omegaBulk1})--(\ref{omegaBulk1b}) and (\ref{OmegaUnif3}) are different by form,
 they both lead to the same result, namely to the thermodynamic potential of the homogeneous system in the random phase approximation.

 The surface contribution to the thermodynamic potential has the form,
 \begin{equation}\label{omegaSurf1}
    \Omega_\surf=\Omega_{0,\surf}+\Delta\Omega_{\surf},
 \end{equation}
 where,
 \begin{equation}\label{omegaSurf1a}
   \frac{\Omega_{0,\surf}}{S}=\frac{\hbar^2p_\mathrm{F}^4}{m\pi^2}
             \left(\frac{\pi}{32}-\frac{d\,p_\mathrm{F}}{15}\right),
 \end{equation}
 is the surface contribution to the thermodynamic potential of non-interacting
 system per unit area (see expression (\ref{omega2IBMsurf})),
 which depends on the Fermi momentum $p_\mathrm{F}$ of interacting electrons,
 \begin{eqnarray}\label{omegaSurf1b}
  \fl  \frac{\Delta\Omega_\surf}{S}=\frac{e^2d\,p_\mathrm{F}^3}{6\pi^2}
      \int\limits_0^\infty\!\dd q
      \Bigg[
       \int\limits_0^1\!\dd \lambda\,\frac{q}{Q(\lambda)}-1
      \Bigg]
      +\frac{e^2p_\mathrm{F}^2}{16\pi}\int\limits_0^\infty\!\dd q\nonumber\\
     \fl +\frac{e^2p_\mathrm{F}^3}{12\pi^2}
      \int\limits_0^\infty\!\dd q\!
      \int\limits_0^1\!\dd \lambda\,
      \frac{q}{Q^2(\lambda)}\frac{Q(\lambda)-q}{Q(\lambda)+q}
       \left[
       1+\frac{3Q^2(\lambda)}{2p_\mathrm{F}^2}
        -\frac{3Q(\lambda)\big(p_\mathrm{F}^2+Q^2(\lambda)\big)}{2p_\mathrm{F}^2}
         \arctan\frac{p_\mathrm{F}}{Q(\lambda)}
      \right]
      \nonumber\\
     \fl-\frac{e^2}{4\pi^4}
      \int\limits_0^\infty\!\dd q\,q\!\!
      \int\limits_0^\infty\!\dd \alpha_1\!\!
      \int\limits_0^\infty\!\dd \alpha_2 \widetilde{J}(q,\alpha_1,\alpha_2)\!
      \int\limits_0^1\!\dd \lambda\,g_2(q,\alpha_1,\alpha_2,\lambda),
 \end{eqnarray}
 \begin{equation*}
   \fl g_2(q,\alpha_1,\alpha_2,\lambda)=\Bigg\{
       2g_1(q,\alpha_1,\alpha_2,\lambda)\,d
       +\left(\frac{Q(\lambda)-q}{Q(\lambda)+q}-2\right)
       \frac{16Q(\lambda)\alpha_1^2\alpha_2^2}{\big[\big(Q^2(\lambda)+\alpha_1^2+\alpha_2^2\big)^2-4\alpha_1^2\alpha_2^2\big]^2}
      \Bigg\}.
 \end{equation*}
 The parameter $d$ for the infinite barrier model is,
 \begin{equation}\label{paramd}
  d=\frac{3\pi}{8p_\mathrm{F}}.
 \end{equation}

 It is worth mentioning that the second term
 in the expression given by Eq. (\ref{omegaSurf1b})
 (it is linear on the chemical potential and is quadratic on the Fermi momentum $p_\mathrm{F}$)
 contains the divergent integral.
 However, as we will see below, this divergent integral disappears in
 the calculation of the internal energy.

 \section{Internal energy}

 \subsection{General expressions}

 By using thermodynamic potential $\Omega$ and
 the Gibbs-Helmholtz equation generalized for the case of variable number of particles,
 we obtain the internal energy of the system~$U$,
 \begin{equation}\label{U}
    U=\Omega-\theta\frac{\partial\Omega}{\partial\theta}-\mu\frac{\partial\Omega}{\partial\mu}.
 \end{equation}
 At low temperatures $\theta\to0$, the second term of the r.h.s. of this equation vanishes and we get,
 \begin{equation}\label{U1}
    U=\Omega+\mu \langle N\rangle,
 \end{equation}
 where we have used the relation
 \begin{equation}\label{eqForChemPot1}
    \langle N\rangle=\frac1\Xi\Sp\left(\ee^{\beta(H-\mu N)}N\right)
                   =-\frac{\partial\Omega}{\partial\mu}.
 \end{equation}

 According to the Eq. (\ref{omega7}), thermodynamic potential can be divided
 into the extensive and surface contributions.
 Then we get,
 \begin{equation}\label{eqForChemPot2}
  \langle N\rangle=-\frac{\partial\big(\Omega_\bulk+\Omega_\surf\big)}{\partial\mu}=N_\bulk+N_\surf,
 \end{equation}
 where,
 \begin{eqnarray}\label{NbulkOmbulk}
    N_\bulk&=-\frac{\partial\Omega_\bulk}{\partial\mu},\\
    N_\surf&=-\frac{\partial\Omega_\surf}{\partial\mu},\label{NsurfOmsurf}
 \end{eqnarray}
 and
 \[
  U=U_\bulk+U_\surf,
 \]
 \begin{eqnarray}
    U_\bulk&=\Omega_\bulk-\mu\frac{\partial\Omega_\bulk}{\partial\mu}=\Omega_\bulk+\mu N_\bulk,\label{Ubulk}\\
    U_\surf&=\Omega_\surf-\mu\frac{\partial\Omega_\surf}{\partial\mu}=\Omega_\surf+\mu N_\surf \label{Usurf}
 \end{eqnarray}
 are the extensive and surface contributions to the internal energy, respectively.

 The chemical potential $\mu$ is the solution of the equation (\ref{eqForChemPot1}).
 By using~(\ref{eqForChemPot2}) in the thermodynamic limit, we get
 \[
  \lim_{N,S,L\to\infty}
  \frac{\langle N\rangle}{SL/2}=
  \lim_{N,S,L\to\infty}
  \frac{N_\bulk}{SL/2}=
  \frac{3}{4\pi}\frac1{r_\mathrm{s}^3},
 \]
 where $r_\mathrm{s}$ is the Wigner-Seitz radius in units of the Bohr radius $a_\mathrm{B}$.
 Moreover, in the thermodynamic limit the summand $N_\surf$ does not affect the chemical potential $\mu$
 (but affects the surface contribution to the internal energy~$U_\surf$) and the equation for $\mu$ can be presented as,
 \begin{equation}\label{eqForChemPot3}
  \langle N\rangle=-\frac{\partial\Omega_\bulk}{\partial\mu}.
 \end{equation}

 According to Eqs. (\ref{Ubulk}) and (\ref{Usurf}),
 in order to calculate the extensive $U_\bulk$
 and the surface $U_\surf$ contributions to the internal energy
 we need to evaluate the  extensive  $N_\bulk$ and surface $N_\surf$
 contributions to the average of the number operator of electrons $\langle N\rangle$.

 \subsection{Average of the number operator of electrons and the chemical potential}

 According to Eqs. (\ref{NbulkOmbulk}) and (\ref{NsurfOmsurf}),
 $N_\bulk$ and $N_\surf$ can be calculated by taking the derivatives
 of $\Omega_\bulk$ and $\Omega_\surf$ with respect to the chemical potential~$\mu$, respectively.
 However, it is easier to use the functional representation of the thermodynamic potential~$\Omega$
 (see (\ref{omegaIntWithoutLambda})) and compute the derivative of this expression (\ref{omega})
 with respect to the chemical potential~$\mu$.
 As a result, we find that,
 \begin{eqnarray}\label{N1}
   \fl \langle N\rangle=\langle N\rangle_0
       +\frac1{2S}\sum\limits_{\mathbf{q}\neq0}\nu(\mathbf{q},0)
        \frac{\partial \langle N\rangle_0}{\partial\mu}
       +\frac1{\beta\Xi_{\mathrm{int}}}\frac{\partial\Xi_{\mathrm{int}}}{\partial\mu}\nonumber\\
   \fl  =\langle N\rangle_0
       +\frac1{2S}\sum\limits_{\mathbf{q}\neq0}\nu(\mathbf{q},0)
        \frac{\partial \langle N\rangle_0}{\partial\mu}\nonumber\\
    \fl +\frac1{\beta\Xi_{\mathrm{int}}}\prod\limits_{\mathbf{q}\neq0}
              \prod\limits_{\nu}\prod\limits_{k}  \big({\textstyle\frac{\beta}{SL}}\nu_k(\mathbf{q})\big)^{-1/2}
              \frac12\!\int\!(\dd \omega)
              J_\mathrm{G}(\omega)\!\sum\limits_{                \begin{array}{c}
                 \scriptstyle \mathbf{q}\neq0,\nu \\[-2mm]
                 \scriptstyle k_1,k_2
                \end{array}} \!\!
              \frac{\partial D_{k_1,k_2}(\mathbf{q},\nu)}{\partial\mu}\,
              \omega_{k_1}(\mathbf{q},\nu)\omega_{k_2}(-\mathbf{q},-\nu)\nonumber\\
   \fl  =\langle N\rangle_0
       +\frac1{2S}\sum\limits_{\mathbf{q}\neq0}\nu(\mathbf{q},0)
       \frac{\partial \langle N\rangle_0}{\partial\mu}+
       \frac1{2\beta}\sum\limits_{                \begin{array}{c}
                 \scriptstyle \mathbf{q}\neq0,\nu \\[-2mm]
                 \scriptstyle k_1,k_2
                \end{array}} \!\!
       \frac{\partial D_{k_1,k_2}(\mathbf{q},\nu)}{\partial\mu}
       \big\langle\omega_{k_1}(\mathbf{q},\nu)\omega_{k_2}(-\mathbf{q},-\nu)\big\rangle_\mathrm{G},
 \end{eqnarray}
 where averaging $\langle\ldots\rangle_\mathrm{G}$ is performed according to Eq. (\ref{bb4}).

 Considering that,
 \begin{equation}\label{averOmegaOmega}
  \big\langle\omega_{k_1}(\mathbf{q},\nu)\omega_{k_2}(-\mathbf{q},-\nu)\big\rangle_\mathrm{G}
  =\frac\beta S g_{k_1,k_2}(\mathbf{q},\nu),
 \end{equation}
 the expression (\ref{N1}) can be rewritten as,
 \begin{equation}\label{N2}
   \fl \langle N\rangle=\langle N\rangle_0
       +\frac1{2S}\sum\limits_{\mathbf{q}\neq0}\nu(\mathbf{q},0)
        \frac{\partial \langle N\rangle_0}{\partial\mu}+\frac1{2S}\sum\limits_{                \begin{array}{c}
                 \scriptstyle \mathbf{q}\neq0,\nu \\[-2mm]
                 \scriptstyle k_1,k_2
                \end{array}} \!\!
     \frac{\partial D_{k_1,k_2}(\mathbf{q},\nu)}{\partial\mu}
     g_{k_1,k_2}(\mathbf{q},\nu).
 \end{equation}
 Taking into account Eqs. (\ref{gFourier}) and (\ref{DFourier}), we obtain,
 \begin{eqnarray}\label{N3}
    \langle N\rangle&=\langle N\rangle_0
       +\frac1{2S}\sum\limits_{\mathbf{q}\neq0}\nu(\mathbf{q},0)
       \frac{\partial \langle N\rangle_0}{\partial\mu}\nonumber\\
       &\quad+\frac1{2SL^2}\sum\limits_{\mathbf{q}\neq0,\nu}
     \int\limits_{-\frac L2}^{+\frac L2}\!\!\dd z_1\!\!
     \int\limits_{-\frac L2}^{+\frac L2}\!\!\dd z_2\;
     \frac{\partial D(\mathbf{q},\nu,z_1,z_2)}{\partial\mu}\,
     g(\mathbf{q},\nu,z_1,z_2),
 \end{eqnarray}
 In order to simplify this expression we make similar approximations,
 as in the calculation of the thermodynamic potential, namely:
 \begin{itemize}
   \item $D\approx{\mathfrak M}^0$;
   \item $g(\mathbf{q},\nu,z_1,z_2)\approx g(\mathbf{q},0,z_1,z_2)\equiv g(\mathbf{q},z_1,z_2)$.
 \end{itemize}
 Then the expression given by Eq. (\ref{N3}) is greatly simplified and,
 \begin{eqnarray}\label{N4}
    \langle N\rangle&=\langle N\rangle_0
       +\frac1{2S}\sum\limits_{\mathbf{q}\neq0}\nu(\mathbf{q},0)
       \frac{\partial \langle N\rangle_0}{\partial\mu}\nonumber\\
       &\quad+\frac1{2SL^2}\sum\limits_{\mathbf{q}\neq0}
     \int\limits_{-\frac L2}^{+\frac L2}\!\!\dd z_1\!\!
     \int\limits_{-\frac L2}^{+\frac L2}\!\!\dd z_2\;
     \sum\limits_{\nu}\frac{\partial {\mathfrak M}^0(\mathbf{q},\nu,z_1,z_2)}{\partial\mu}\,
     g(\mathbf{q},z_1,z_2),
 \end{eqnarray}
 where the summation by the frequency $\nu$  applies only to the
 derivative of the two-particle correlator with respect to the chemical potential~$\mu$
 and can be performed analytically,
 \begin{eqnarray}\label{sumPoNuPohidnaM0}
 \fl \sum\limits_{\nu}\frac{\partial{\mathfrak M}^0(\mathbf{q},\nu,z_1,z_2)}{\partial\mu}=-L^2
  \sum_{{\bf p},\alpha_1}\frac{\partial n_{\alpha_1}({\bf p})}{\partial\mu}
  |\varphi_{\alpha_1}\!(z_1)|^2 \dirac(z_1-z_2)\nonumber\\
  \fl\quad +L^2\sum_{{\bf p},\alpha_1,\alpha_2}\frac{\partial n_{\alpha_1}({\bf p})}{\partial\mu}
    \big(n_{\alpha_2}({\bf p}-\mathbf{q})+n_{\alpha_2}({\bf p}+\mathbf{q})\big)
  \varphi^*_{\alpha_1}\!(z_1) \varphi^{\vphantom{*}}_{\alpha_2}\!(z_1)
      \varphi^*_{\alpha_2}\!(z_2) \varphi^{\vphantom{*}}_{\alpha_1}\!(z_2).
 \end{eqnarray}

 Substituting Eq. (\ref{sumPoNuPohidnaM0}) into Eq. (\ref{N4}), we find,
 \begin{eqnarray}\label{N5}
   \fl \langle N\rangle=\langle N\rangle_0
       +\frac1{2S}\sum\limits_{\mathbf{q}\neq0}\nu(\mathbf{q},0)
       \frac{\partial \langle N\rangle_0}{\partial\mu}-
       \frac1{2S}\sum\limits_{\mathbf{q}\neq0}\sum\limits_{\mathbf{p},\alpha}
           \frac{\partial n_\alpha(\mathbf{p})}{\partial\mu}
      \int\limits_{-\frac L2}^{+\frac L2}\!\!\dd z\,|\varphi_\alpha(z)|^2
       g(\mathbf{q},z,z)\\
    \fl+\frac1{2S}\sum\limits_{\mathbf{q}\neq0}\sum\limits_{\mathbf{p},\alpha_1,\alpha_2}
      \frac{\partial\big(n_{\alpha_1}(\mathbf{p}) n_{\alpha_2}(\mathbf{p}-\mathbf{q})\big)}{\partial\mu}
                     \int\limits_{-\frac L2}^{+\frac L2}\!\!\dd z_1\!\!
      \int\limits_{-\frac L2}^{+\frac L2}\!\!\dd z_2\,
      \varphi^*_{\alpha_1}\!(z_1) \varphi^{\vphantom{*}}_{\alpha_2}\!(z_1)
      \varphi^*_{\alpha_2}\!(z_2) \varphi^{\vphantom{*}}_{\alpha_1}\!(z_2)
      g(\mathbf{q},z_1,z_2).\nonumber
 \end{eqnarray}

 Taking into account the expressions for the one- (\ref{unarna}) and two-particle (\ref{binarna})
 distribution functions of electrons in the semi-infinite jellium \cite{JPS2003_2},
 we get,
 \begin{eqnarray}\label{N6}
    \fl\langle N\rangle=\langle N\rangle_0
       +\frac1{2S}\sum\limits_{\mathbf{q}\neq0}\nu(\mathbf{q},0)
        \frac{\partial \langle N\rangle_0}{\partial\mu}-
        \frac1{2}\frac{S}{V}\int\limits_{-\frac L2}^{+\frac L2}\!\!\dd z\,
      \frac{\partial \left(\langle N\rangle_0F_1^0(z)\right)}{\partial\mu}\,
      g(\mathbf{r}_{||},z,z)\big|_{\mathbf{r}_{||}=0}\nonumber\\
   \fl -\frac1{2}\frac{S}{V^2}
      \int\limits_{S}\!\dd \mathbf{r}_{||}\!\!
      \int\limits_{-\frac L2}^{+\frac L2}\!\!\dd z_1\!\!
      \int\limits_{-\frac L2}^{+\frac L2}\!\!\dd z_2\,
      \frac{\partial \left(
        \langle N\rangle_0^2\big(F_2^0(\mathbf{r}_{||},z_1,z_2)-F_1^0(z_1)F_1^0(z_2)\big)
      \right)}{\partial\mu} g(\mathbf{r}_{||},z_1,z_2),
 \end{eqnarray}
 where $g(\mathbf{r}_{||},z_1,z_2)$
 is the effective inter-electron interaction in the coordinate representation
 (note that the transition to the coordinate representation is similar to Eq. (\ref{gKoord})).

\subsection{Calculation of $\langle N\rangle_0$ and ${\partial\langle N\rangle_0}/{\partial\mu}$}

 By using the thermodynamic potential of the non-interacting system $\Omega_0$
 (see Eq. (\ref{Omega0Suma})), we calculate $\langle N\rangle_0$ at low temperatures,
 \[
  \langle N\rangle_0=-\frac{\partial\Omega_0}{\partial\mu}
                    =-\frac{\partial\big(\Omega_{0,\bulk}+\Omega_{0,\surf}\big)}{\partial\mu}
                    =N_{0,\bulk}+N_{0,\surf},
 \]
 where,
 \begin{equation}\label{N0bulk}
     N_{0,\bulk}=-\frac{\partial\Omega_{0,\bulk}}{\partial\mu}
                =\frac{SL}{2}\frac{2\sqrt{2}m^{3/2}}{3\pi^2\hbar^3}\mu^{3/2}=\frac{SL}{2}\frac{p_\mathrm{F}^3}{3\pi^2}
 \end{equation}
 is the extensive contribution $\langle N\rangle_0$, and
 \begin{equation}\label{N0surfIBM}
   \fl  N_{0,\surf}=-\frac{\partial\Omega_{0,\surf}}{\partial\mu}=
      S\left(\frac{2\sqrt{2}m^{3/2}d}{3\pi^2\hbar^3}\mu^{3/2}
                   -\frac{m}{4\pi\hbar^2}\mu\right)=
      S\,\frac{p_\mathrm{F}^2}{\pi^2}\left(\frac{d\,p_\mathrm{F}}{3}-\frac{\pi}{8}\right)
 \end{equation}
 is the surface contribution $\langle N\rangle_0$.
 The latter result reduces to the one obtained previously in Refs. \cite{Sugiyama,Sugiyama2},
 if we put $d=0$ in Eq. (\ref{N0surfIBM}).

 By using $\langle N\rangle_0$,
 we calculate ${\partial\langle N\rangle_0}/{\partial\mu}$
 \[
  \frac{\partial\langle N\rangle_0}{\partial\mu}=
  \frac{\partial N_{0,\bulk}}{\partial\mu}+
  \frac{\partial N_{0,\surf}}{\partial\mu},
 \]
 where the extensive contribution is,
 \begin{equation}\label{pohN0bulk}
     \frac{\partial N_{0,\bulk}}{\partial\mu}=
     \frac{SL}{2}\frac{\sqrt{2}m^{3/2}}{\pi^2\hbar^3}\mu^{1/2}
     =\frac{SL}{2}\frac{m}{\hbar^2}\frac{p_\mathrm{F}}{\pi^2},
 \end{equation}
 and the surface contribution is,
 \begin{equation}\label{pohN0surfIBM}
    \frac{\partial N_{0,\surf}}{\partial\mu}=S\,\left(\frac{\sqrt2 m^{3/2}d}{\pi^2\hbar^3}\mu^{1/2}
     -\frac{m}{4\pi \hbar^2}\right) =S\frac{m}{\hbar^2\pi^2}\left({d\,p_\mathrm{F}}-\frac{\pi}{4}\right).
 \end{equation}

 It should be noted,
 that the following relations are valid for the non-interacting homogeneous system,
 \begin{eqnarray}\label{N0unif}
     N_{0,\unif}&=V\frac{p_\mathrm{F}^3}{3\pi^2},
 \end{eqnarray}
 \begin{eqnarray}
    \frac{\partial N_{0,\unif}}{\partial\mu}&=
    V\frac{m}{\hbar^2}\frac{p_\mathrm{F}}{\pi^2},\label{pohN0unif}.
 \end{eqnarray}
 They coincide with the extensive contributions given by Eqs. (\ref{N0bulk}) and (\ref{pohN0bulk}),
 respectively.

\subsubsection*{Average of the number operator of electrons for the infinite barrier model.}

 For further calculation of (\ref{N5}) or (\ref{N6}),
 as in above,
 we consider the infinite barrier model (\ref{poverhPot}) of the surface potential $V_\surf(z)$.
 After the summation by momenta $\mathbf{p}$ in Eq. (\ref{N5}),
 we get,
 \begin{equation}\label{Nbulk1}
    N_\bulk=N_{0,\bulk}+\Delta N_\bulk,
 \end{equation}
 where,
 \begin{equation}\label{Nbulk1a}
    \frac{N_{0,\bulk}}{SL/2}=\frac{p_\mathrm{F}^3}{3\pi^2}
 \end{equation}
 is the extensive contribution to the average of the number operator $\langle N\rangle$
 of the non-interacting system per unit volume (see expression (\ref{N0bulk})).
 This contribution depends on the Fermi momentum $p_{\mathrm{F}}$ of the interacting electrons.
 $\Delta N_\bulk$ has the form,
 \begin{eqnarray}\label{Nbulk1b}
    \frac{\Delta N_\bulk}{SL/2}&=\frac{p_\mathrm{F}}{2\pi^2a_\mathrm{B}}
             \int\limits_0^\infty\!\dd q
             \left(1-\frac{q}{Q}\right)\nonumber\\
             &\quad+\frac{2}{\pi^4a_\mathrm{B}}
             \int\limits_0^\infty\!\dd q \, q
             \int\limits_0^\infty\!\dd \alpha_1 \!
             \int\limits_0^\infty\!\dd \alpha_2\,
             I_-(q,\alpha_1,\alpha_2)\,
             g_1(q,\alpha_1,\alpha_2,1),
 \end{eqnarray}
 ${Q\equiv Q(1)}$,
 \begin{equation}\label{I21}
   \fl I_-(q,\alpha_1,\alpha_2)=
    \left\{
      \begin{array}{cl}
        \!\!\left\{
         \begin{array}{ll}
            0, & 0\leqslant q\leqslant c_1-c_2, \\
            \arccos\frac{q^2+c_1^2-c_2^2}{2c_1q}, & c_1-c_2<q\leqslant c_1+c_2,\\
            0, & c_1+c_2<q<\infty,
         \end{array}\!\!\!\!
        \right\}, & \!\!\! c_1>c_2,\\[8mm]
        \!\!\left\{
         \begin{array}{ll}
            \pi, & 0\leqslant q\leqslant c_2-c_1, \\
            \arccos\frac{q^2+c_1^2-c_2^2}{2c_1q}, & c_2-c_1<q\leqslant c_1+c_2,\\
            0, & c_1+c_2<q<\infty,
         \end{array}\!\!\!\!
        \right\}, & \!\!\! c_2\geqslant c_1,
      \end{array}
    \right.
 \end{equation}
 where,
 \[
  c_1=\sqrt{p_\mathrm{F}^2-\alpha_1^2},\quad
  c_2=\sqrt{p_\mathrm{F}^2-\alpha_2^2}.
 \]

 It should be noted that
 the dividing plane has no effect on the expressions (\ref{Nbulk1})--(\ref{Nbulk1b}) and therefore
 Eq. (\ref{Nbulk1}) is the number of electrons of the homogeneous system per unit volume.
 However, the number of electrons of the homogeneous system can similarly be calculated
 from the Hamiltonian of the infinite jellium (\ref{HamiltonianUnif}).
 These calculations are much simpler and with similar approximations we obtain,
 \begin{equation}\label{Nunif2}
   \fl \frac{N_\unif}{V}=\frac{p_\mathrm{F}^3}{3\pi^2}
             +\frac{p_\mathrm{F}}{4\pi^4}
              \frac{m}{\hbar^2}
             \int\limits_0^\infty\!\dd q\,q^2\,\big(\nu(q)-g_\unif(q)\big)
            +\frac{p_\mathrm{F}}{4\pi^4}\frac{m}{\hbar^2}
             \int\limits_0^\infty\!\dd q\,q^2\,
             {\widetilde{I}^-_\unif(q)}\,g_\unif(q),
 \end{equation}
 where,
 \begin{equation*}
     \widetilde{I}^-_\unif(q)=\left\{
      \begin{array}{cl}
        2, & \frac{q}{2p_\mathrm{F}}\leqslant-1, \\
        1-\frac{q}{2p_\mathrm{F}}, & -1<\frac{q}{2p_\mathrm{F}}\leqslant1, \\
        0, & \frac{q}{2p_\mathrm{F}}>1 ,
      \end{array}
      \right.
\end{equation*}
 \[
  g_\unif(q)=\frac{4\pi e^2}{q^2+\varkappa_{\mathrm{TF}}^2\lindhard\big(\textstyle\frac{q}{2p_\mathrm{F}}\big)}
 \]
 is the three-dimensional Fourier-transform of the effective inter-electron interaction of the homogeneous system.

 Substituting the expressions for the three-dimensional Fourier-transforms of the Coulomb interaction $\nu(q)$
 and the effective inter-electron interaction $g_\unif (q) $ into Eq. (\ref{Nunif2}), we get,
 \begin{equation}\label{Nunif3}
    \frac{N_\unif}{V}=\frac{p_\mathrm{F}^3}{3\pi^2}
             +\frac{p_\mathrm{F}}{\pi^3a_\mathrm{B}}
             \int\limits_0^\infty\!\!\dd q\,
             \frac{\big(
              \varkappa_{\mathrm{TF}}^2\lindhard\big({\textstyle\frac{q}{2p_\mathrm{F}}}\big)
              +q^2\widetilde{I}^-_\unif(q)
             \big)}
                  {q^2+\varkappa_{\mathrm{TF}}^2\lindhard\big(\textstyle\frac{q}{2p_\mathrm{F}}\big)}.
 \end{equation}

 The expressions (\ref{Nbulk1})--(\ref{Nbulk1b}) and (\ref{Nunif3})
 are nonlinear algebraic equations for the chemical potential~$\mu$ ($\mu=\frac{\hbar^2p_\mathrm{F}^2}{2m}$).
 Although their form is different but they yield the same result,
 because the dividing plane does not affect the chemical potential.

\begin{figure}[hbtp]
  \centering
  \includegraphics[width=0.7\textwidth]{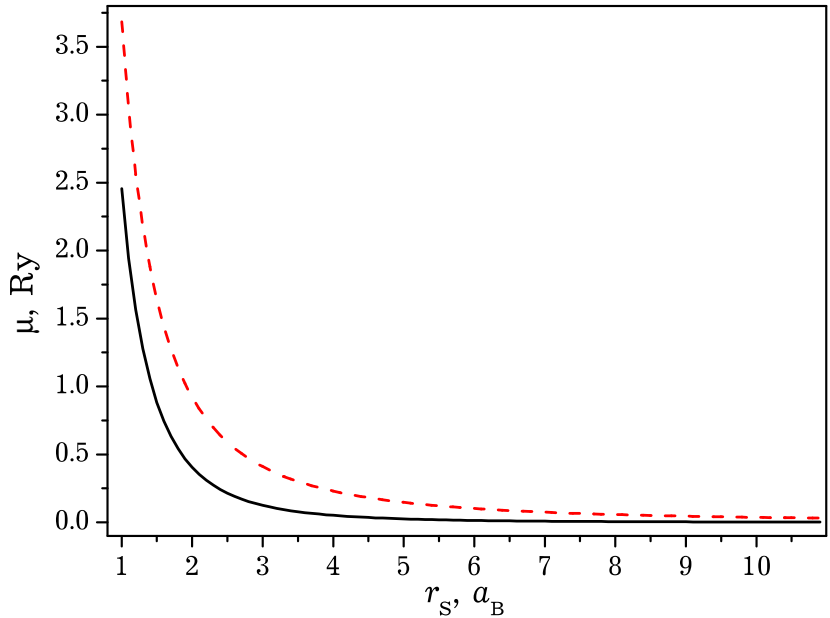}\\
  \caption{The chemical potential of the interacting electrons~$\mu$ (solid line) and the chemical potential
  of the non-interacting electrons~$\mu_0$ (dashed line) as a function of the Wigner-Seitz radius.}\label{chemPot}
\end{figure}

 In Figure\,\ref{chemPot} the chemical potential
 as the function of the Wigner-Seitz radius $r_\mathrm{s}$ is  presented.
 The nomenclature of lines is given in the figure caption.
 It is worth noting that the chemical potential is the solution of nonlinear algebraic equations (\ref{Nunif3}),
 and the chemical potential of non-interacting electrons is,
 \begin{equation}\label{mu0}
  \mu_0=\left(\frac{9\pi}{4}\right)^{2/3}\frac1{(r_\mathrm{s}/a_\mathrm{B})^2},\; \mathrm{Ry}.
 \end{equation}

 It can be seen that taking into account the Coulomb interaction
 leads to a decrease of the chemical potential of electrons.

\begin{figure}[hbtp]
  \centering
  \includegraphics[width=0.7\textwidth]{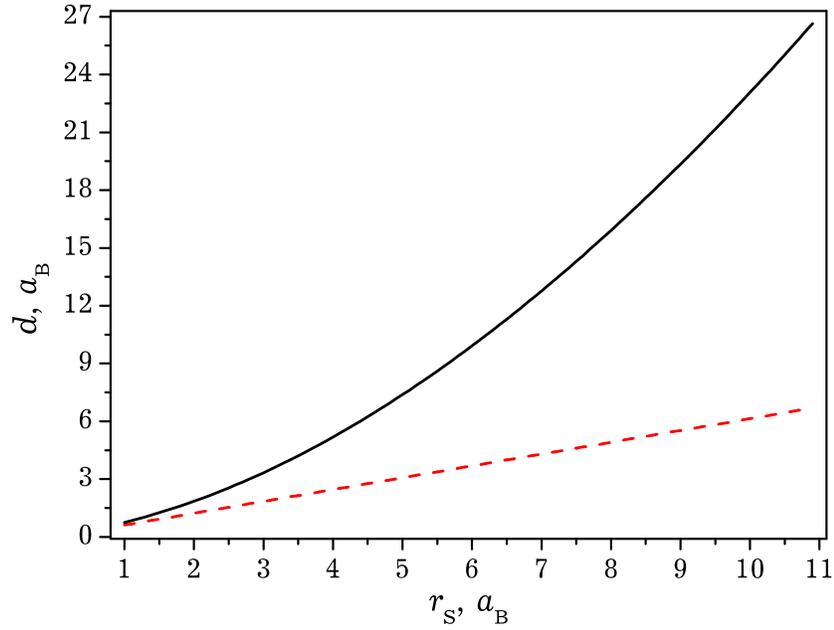}\\
  \caption{The parameter $d$ as a function of Wigner-Seitz radius (solid line is for interacting electrons whereas the dashed line is for non-interacting electrons).}\label{FIGd}
\end{figure}

 In Figure\,\ref{FIGd} the parameter, $d$, $d=3\pi/(8p_\mathrm{F})$
 as a function of Wigner-Seitz radius $r_\mathrm{s}$ is given.
 The parameter $d$ is the distance from the infinite potential barrier to the dividing plane ($z=0$).
 The conclusion from the curves is that taking into account
 the Coulomb interaction between electrons leads to
 an increase of this distance and its nonlinear dependence on $r_\mathrm{s}$,
 whereas the parameter $d$ for the non-interacting system is a linear function of $r_\mathrm{s}$.

\begin{figure}[hbtp]
  \centering
  \includegraphics[width=0.7\textwidth]{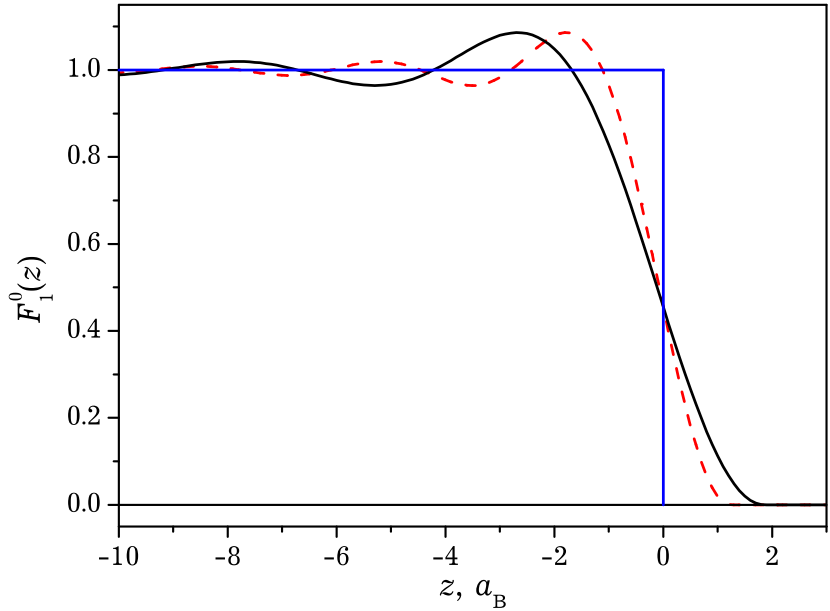}\\
  \caption{The one-particle distribution function of electrons as a function of the  electron coordinate normal to the dividing plane at $r_\mathrm{s}=2\,a_\mathrm{B}$ (the solid line is for interacting electrons whereas the dashed line is for non-interacting electrons).}\label{unarna2}
\end{figure}
\begin{figure}[hbtp]
  \centering
  \includegraphics[width=0.7\textwidth]{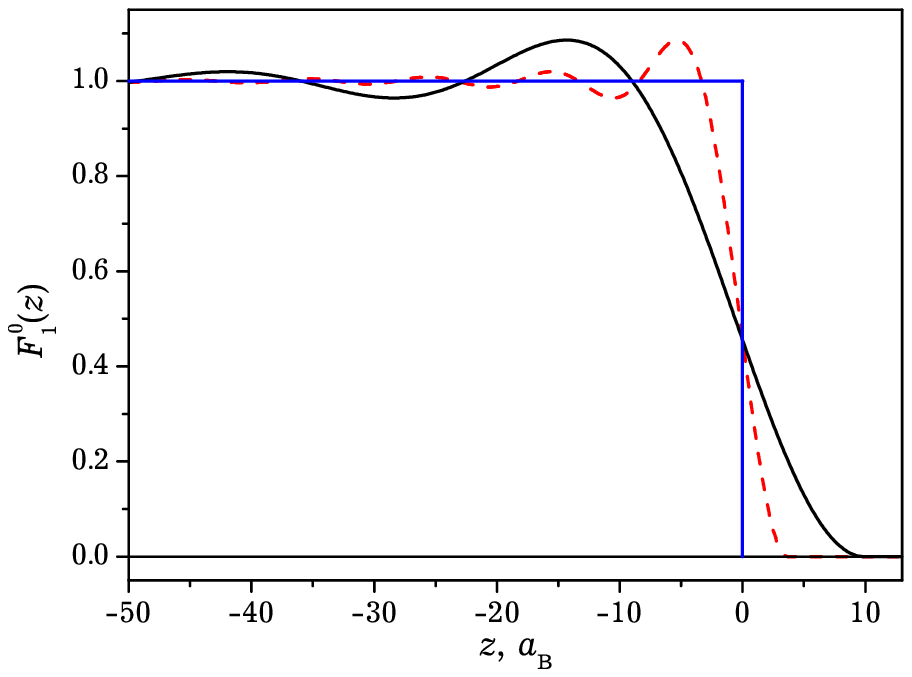}\\
  \caption{The one-particle distribution function of electrons as a function of the  electron coordinate normal to the dividing plane at $r_\mathrm{s}=6\,a_\mathrm{B}$ (the solid line is for interacting electrons whereas the dashed line is for non-interacting electrons).}\label{unarna6}
\end{figure}

 In Figures\,\ref{unarna2} and \ref{unarna6},
 the one-particle distribution function of electrons (\ref{unarnaIBM}) is presented for the following values
 of Wigner-Seitz radius: ${r_\mathrm{s}=2\,a_\mathrm {B}}$ and ${r_\mathrm {s}=6\,a_\mathrm{B}}$.
 The one-particle function of electrons (solid lines)
 depends on the chemical potential which is the solution of nonlinear algebraic equations (\ref{Nunif3})
 (here the Coulomb interaction is taken into account).
 The positive charge is located at $z\leqslant0$.
 It can be concluded that taking into account the Coulomb interaction
 leads to an increase of period of damped oscillations
 of the one-particle distribution function around its value in the body of the metal
 which equals to unity.

 The surface contribution to the average number of electrons $\langle N\rangle$ has the form,
 \begin{equation}\label{Nsurf1}
     N_\surf=N_{0,\surf}+\Delta N_\surf,
 \end{equation}
 where,
 \begin{equation}\label{Nsurf1a}
     \frac{N_{0,\surf}}S=\frac{p_\mathrm{F}^2}{\pi^2}
                           \left(\frac{d\,p_\mathrm{F}}{3}-\frac{\pi}8\right),
 \end{equation}
 is the surface contribution to the average number of the non-interacting electrons system
 per unit area (see Eq. (\ref{N0surfIBM})),
 that depends on the Fermi momentum $p_\mathrm{F}$ of the interacting electrons, and
 \begin{eqnarray}\label{Nsurf1b}
    \frac{\Delta N_\surf}S&=\frac{d\,p_\mathrm{F}}{2\pi^2a_\mathrm{B}}
             \int\limits_0^\infty\!\dd q
             \left(1-\frac{q}{Q}\right)
             -\frac1{8\pi a_\mathrm{B}}\int\limits_0^\infty\!\dd q \nonumber\\
           &\quad  -\frac{p_\mathrm{F}}{4\pi^2a_\mathrm{B}}
              \int\limits_0^\infty\!\dd q\,\frac{q}{Q^2}\,
              \frac{Q-q}{Q+q}
              \left(1-\frac{Q}{p_\mathrm{F}}\arctan\frac{p_\mathrm{F}}{Q}\right)  \nonumber\\
            &\quad+\frac{1}{\pi^4a_\mathrm{B}}
             \int\limits_0^\infty\!\dd q \, q
             \int\limits_0^\infty\!\dd \alpha_1 \!
             \int\limits_0^\infty\!\dd \alpha_2\,
             I_-(q,\alpha_1,\alpha_2)\,
             g_1(q,\alpha_1,\alpha_2,1).
 \end{eqnarray}

 It should be noted that the second term in Eq. (\ref{Nsurf1b}),
 (which is linear on the chemical potential and is quadratic on the Fermi momentum $p_\mathrm{F}$)
 contains the divergent integral.
 However, as we will see below, this divergent integral disappears in
 the calculation of the internal energy.

 Also it should be noted that
 if we substitute the value of $d$~(\ref{paramd})
 into the expression (\ref{Nsurf1a}),  
 it vanishes.
 If we take into account the Coulomb interaction between electrons,
 then ${N_{0,\surf}\neq0}$.

 \subsection{Internal energy and surface energy}

 Substituting Eqs. (\ref{omega6}) and (\ref{N6})
 into Eq.~(\ref{U1}),
 we obtain the internal energy.
 This expression can be considered as one possible energy functional
 that in contrast to the functionals used in the density functional theory,
 depends not only on the one-particle distribution function of the electrons,
 but also on the two-particle distribution function and the effective inter-electron interaction.

 The extensive contribution to the internal energy per unit volume
 follows from the substitution of $\Omega_\bulk$ and $N_\bulk$
 (or $\Omega_\unif$ and $N_\unif$) into Eq. (\ref{Ubulk}).
 Then we get,
 \begin{eqnarray}\label{Ubulk3}
    \frac{U_\bulk}{SL/2}&=\frac{\hbar^2}{10m\pi^2}p_\mathrm{F}^5
     -\frac{e^2p_\mathrm{F}^3}{6\pi^2}
      \int\limits_0^\infty\!\dd q
      \Bigg[
       1-\int\limits_0^1\!\dd \lambda\,\frac{q}{Q(\lambda)}
      \Bigg]+\frac{e^2p_\mathrm{F}^3}{4\pi^2}
             \int\limits_0^\infty\!\dd q
             \left(1-\frac{q}{Q}\right)\nonumber\\
     &\quad+\frac{e^2p_\mathrm{F}^2}{\pi^4}
             \int\limits_0^\infty\!\dd q \, q
             \int\limits_0^\infty\!\dd \alpha_1 \!
             \int\limits_0^\infty\!\dd \alpha_2\,
             I_-(q,\alpha_1,\alpha_2)\,
             g_1(q,\alpha_1,\alpha_2,1) \nonumber\\
     &\quad-\frac{e^2}{2\pi^4}
      \int\limits_0^\infty\!\dd q\,q\!
      \int\limits_0^\infty\!\dd \alpha_1\!\!
      \int\limits_0^\infty\!\dd \alpha_2
      \widetilde{J}(q,\alpha_1,\alpha_2)
      \int\limits_0^1\!\dd \lambda\,\,
      g_1(q,\alpha_1,\alpha_2,\lambda),
 \end{eqnarray}
 or
 \begin{eqnarray}\label{Uunif3}
   \fl \frac{U_\unif}{V}=\frac{\hbar^2}{10m\pi^2}p_\mathrm{F}^5\\
    \fl +\frac{e^2p_\mathrm{F}^3}{\pi^3}
      \int\limits_0^\infty\!\dd q\,
      \Bigg[
       \frac12\frac{\big(
              \varkappa_{\mathrm{TF}}^2\lindhard\big({\textstyle\frac{q}{2p_\mathrm{F}}}\big)
              +q^2\widetilde{I}^-_\unif(q)
             \big)}
                  {q^2+\varkappa_{\mathrm{TF}}^2\lindhard\big(\textstyle\frac{q}{2p_\mathrm{F}}\big)}
      -\frac13+
       \frac{q^2\big(1-\widetilde{J}_\unif(q)\big)}{3\varkappa^2_{\mathrm{TF}}L\big(\textstyle\frac{q}{2p_\mathrm{F}}\big)}
       \ln\left(1+\frac{\varkappa^2_{\mathrm{TF}}}{q^2}L
       \big(\textstyle\frac{q}{2p_\mathrm{F}}\big)\right)
      \Bigg],\nonumber
 \end{eqnarray}
 where the first term of the expression~(\ref{Ubulk3}) (or~(\ref{Uunif3}))
 is the internal energy of non-interacting system.

 By using the expressions for $\Omega_\surf$ and $N_\surf$ in Eq. (\ref{Usurf})
 we obtain the surface contribution to the internal energy per unit area.
 We are interested in the case of low temperatures ($\theta\to0$).
 Then according to \cite{Kiejna,Paash},
 the ratio ${{U_\surf}/{S}}$ is the free surface energy~$\sigma $,
 and the magnitude of ${{U_\surf}=\sigma S}$ is the work
 that is necessary for irreversible process of creating a new free surface~$S$.
 The quantity $\sigma$ describes excess energy of surface area compared with the energy inside the body of the metal.
 Then, the surface energy~$\sigma$ can be presented as,
 \begin{equation}\label{surfEn}
    \sigma=\sigma_0+\Delta\sigma,
 \end{equation}
 where,
 \begin{equation}\label{surfEn0}
    \sigma_0=\frac{\hbar^2p_\mathrm{F}^4}{2m\pi^2}
             \left(\frac{d\,p_\mathrm{F}}{5}-\frac{\pi}{16}\right),
 \end{equation}
 \begin{eqnarray}\label{surfEn1}
   \fl \Delta\sigma=\frac{e^2d\,p_\mathrm{F}^3}{4\pi^2}\int\limits_0^\infty\!\dd q\left(1-\frac{q}{Q}\right)
      -\frac{e^2d\,p_\mathrm{F}^3}{6\pi^2}\int\limits_0^\infty\!\dd q
      \Bigg[1-\int\limits_0^1\!\dd \lambda\,\frac{q}{Q(\lambda)}\Bigg]\nonumber\\
   \fl  -\frac{e^2p_\mathrm{F}^3}{8\pi^2}\int\limits_0^\infty\!\dd q\,\frac{q}{Q^2}\,
              \frac{Q-q}{Q+q}\left(1-\frac{Q}{p_\mathrm{F}}\arctan\frac{p_\mathrm{F}}{Q}\right) \nonumber\\
   \fl  +\frac{e^2p_\mathrm{F}^3}{12\pi^2}
      \int\limits_0^\infty\!\dd q\!
      \int\limits_0^1\!\dd \lambda\,
      \frac{q}{Q^2(\lambda)}\frac{Q(\lambda)-q}{Q(\lambda)+q}
       \left[
       1+\frac{3Q^2(\lambda)}{2p_\mathrm{F}^2}
        -\frac{3Q(\lambda)\big(p_\mathrm{F}^2+Q^2(\lambda)\big)}{2p_\mathrm{F}^2}
         \arctan\frac{p_\mathrm{F}}{Q(\lambda)}
      \right]\nonumber\\
   \fl  +\frac{e^2p_\mathrm{F}^2}{2\pi^4}
             \int\limits_0^\infty\!\dd q \, q
             \int\limits_0^\infty\!\dd \alpha_1 \!
             \int\limits_0^\infty\!\dd \alpha_2\,
             I_-(q,\alpha_1,\alpha_2)\,
             g_1(q,\alpha_1,\alpha_2,1) \nonumber\\
   \fl  -\frac{e^2}{4\pi^4}
      \int\limits_0^\infty\!\dd q\,q\!
      \int\limits_0^\infty\!\dd \alpha_1\!\!
      \int\limits_0^\infty\!\dd \alpha_2
      \widetilde{J}(q,\alpha_1,\alpha_2)
      \int\limits_0^1\!\!\dd \lambda\,\,
      g_2(q,\alpha_1,\alpha_2,\lambda).
 \end{eqnarray}

 By placing the parameter $d$~(\ref{paramd}) in Eq.~(\ref{surfEn0}), we obtain,
 \begin{equation}\label{surfEn01}
    \sigma_0=\frac{\hbar^2p_\mathrm{F}^4}{160m\pi}.
 \end{equation}
 It coincides by form with the surface energy of non-interacting system
 \cite{Huntington,Stratton, Moore1976, Sugiyama, Sugiyama2},
 but in Eq. (\ref{surfEn01}), $p_\mathrm{F}$ is the Fermi momentum
 that takes into account the Coulomb interaction between electrons.

\begin{figure}[hbtp]
  \centering
  \includegraphics[width=0.7\textwidth]{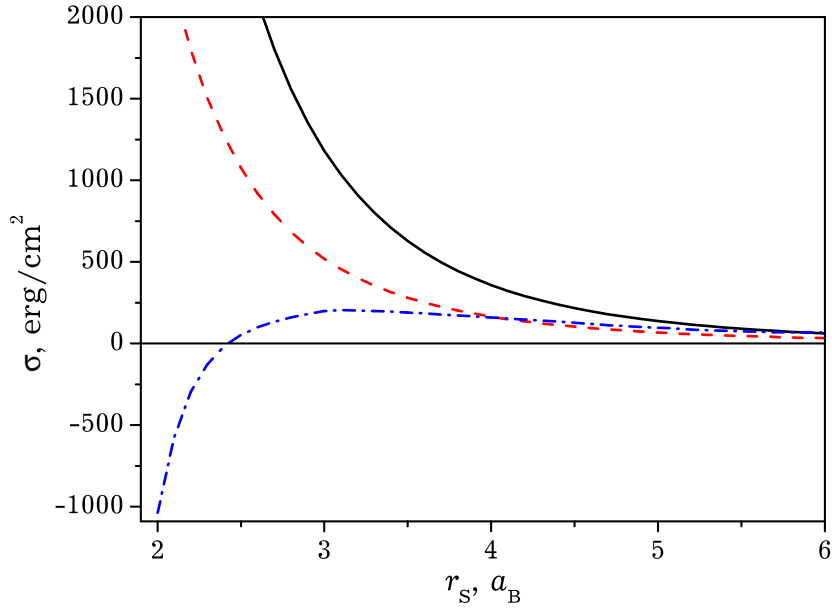}\\
  \caption{The surface energy as a function of Wigner-Seitz radius
           (the solid line is for interacting system,
           the dashed line is for noninteracting system whereas the
           dash-dotted line is the result of Lang and Kohn \cite{Lang}).}\label{surfaceEnergy}
\end{figure}

 In Figure\,\ref{surfaceEnergy}
 the dependence of the surface energy on Wigner-Seitz radius $r_\mathrm{s}$ is presented.
 The solid line is the surface energy calculated using the formulas (\ref{surfEn}), (\ref{surfEn1}), (\ref{surfEn01})
 the solution of nonlinear algebraic equation (\ref{Nunif3}) yields the chemical potential.
 The dashed line is the surface energy of non-interacting system (\ref{surfEn01}),
 the dash-dotted line is the result of Lang and Kohn \cite{Lang}.

 The result given in this figure show that taking into account the Coulomb interaction
 leads to an increase of the surface energy compared to the surface energy of non-interacting system.
 In addition,
 the surface energy calculated by us is positive in the entire region of $r_\mathrm{s}$
 and in the interval ${r_\mathrm{s}=5{.}5-6\,a_\mathrm {B}}$ it coincides with the surface energy calculated by Lang and Kohn.

 \section{Conclusions}

 The general expression for the thermodynamic potential of the semi-infinite jellium model is obtained
 by using the method of functional integration.
 The knowledge of the two-particle correlation function of electrons and
 the effective inter-electron is required for the practical application of this expression.

 It is shown that taking into account the Coulomb interaction between electrons leads to a decrease of the chemical potential.
 It is also shown that  the surface contribution to the  thermodynamic potential does not affect the chemical potential.

 By using the infinite barrier model, the extensive and surface contributions to the thermodynamic potential,
 the average of the number operator of electrons and the internal energy are obtained and studied at low temperatures.

 The influence of the Coulomb interactions between electrons on the behavior of the one-particle distribution function is studied as well.
 We obtained that taking into account the Coulomb interaction between electrons leads to
 an increase of the period of damped oscillations around its average value in the body of the metal.

 It is found that
 taking into account the Coulomb interaction between electrons,
 the distance between the dividing plane and the surface potential
 as a function of the Wigner-Seitz radius looses its linear behavior.
 Namely, it grows faster.

 Based on the expression for the surface contribution to the internal energy
 and modeling the surface potential by the infinite barrier,
 the surface energy is calculated at low temperatures.
 It is shown that taking into account the Coulomb interaction between electrons results in a growth of the surface energy.
 The surface energy is positive in the entire range of Wigner-Seitz radius.
 In the interval ${r_\mathrm{s}>5\,a_\mathrm{B}}$, the surface energy calculated by us is in a good agreement with calculations of Lang and Kohn~\cite{Lang}.

 Our calculations of the chemical potential and surface energy demonstrate
 that taking into account the Coulomb interaction between electrons is very important in the region of small ${r_\mathrm {s}}$,
 and that the influence of Coulomb interaction on these values decreases with increasing ${r_\mathrm{s}}$.

 \appendix
 \renewcommand{\theequation}{\Alph{section}.\arabic{equation}}

 \section{Density of states}\label{DOSapp}

 Let us calculate the density of states of electrons without Coulomb interaction,
 \begin{equation}\label{DOS0}
  \rho(E)=\sum\limits_{\mathbf{p},\alpha}
  \dirac\big(E-E_\alpha(\mathbf{p})\big),
 \end{equation}
 where according to (\ref{3.2})
 \[
  E_\alpha(\mathbf{p})=\frac{\hbar^2(p^2+\alpha^2)}{2m}.
 \]

 In the thermodynamic limit (${S\to\infty}$ and ${L\to\infty}$,
 the sum can be replaced by the integral,
 according to the Euler-Maclaurin formula \cite{Vakarchuk,Paash3},
 \begin{equation*}
   \fl \sum\limits_{n=0}^\infty f(n)=\int\limits_0^\infty\!\!f(x)\,\dd x
      -\frac12\big[f(\infty)-f(0)\big]
      +\frac{B_1}{2!}\big[f'(\infty)-f'(0)\big]-
      \frac{B_2}{4!}\big[f'''(\infty)-f'''(0)\big]+\ldots,
 \end{equation*}
 where $B_k$ are the Bernoulli numbers.

 We perform summation over two-dimensional vector
 according to Eq. (\ref{impulsP}) and obtain,
 \begin{equation}\label{sumToIntPoP}
   \fl \sum\limits_{\mathbf{p}}f(\mathbf{p})
    =2\int\limits_{-\infty}^{+\infty}\!\!\!\dd n_x\!\!
      \int\limits_{-\infty}^{+\infty}\!\!\!\dd n_y f(\mathbf{p})
    =\frac{2S}{(2\pi)^2}\!\!
      \int\limits_{-\infty}^{+\infty}\!\!\!\dd p_x\!\!
      \int\limits_{-\infty}^{+\infty}\!\!\!\dd p_y \,f(\mathbf{p})=\frac{2S}{(2\pi)^2}\!\!
      \int\limits_{-\infty}^{+\infty}\!\!\!\dd \mathbf{p}\,f(\mathbf{p}),
 \end{equation}
 where two possible orientations of the electron spin is taken into account.

 The summation over three-dimensional vector $\mathbf{p}$ yields,
 \begin{equation}\label{sumToIntPoPdim3}
    \sum\limits_{\mathbf{p}}f(\mathbf{p})=\frac{2V}{(2\pi)^3}\!\!
      \int\limits_{-\infty}^{+\infty}\!\!\!\dd \mathbf{p}\,f(\mathbf{p}).
 \end{equation}

 Now we consider the summation over $\alpha$.
 Eq. (\ref{eqForAlfa}) implies that,
 \[
  \frac{\dd n}{\dd\alpha}=
  \frac{L}{2\pi}
  \left(
   1+\frac{2d}L
  \right),
 \]
 then,
 \begin{eqnarray}\label{sumToIntPoAlfa}
    \sum\limits_{\alpha}f(\alpha)
    &=\int\limits_{0}^{+\infty}\!\!\!\dd n\,f(\alpha)-\frac12f(0)
     =\int\limits_{0}^{+\infty}\!\frac{\dd n}{\dd\alpha}\,\dd \alpha f(\alpha)-\frac12f(0)\nonumber\\
    &=\int\limits_{0}^{+\infty}\!\!\!\dd \alpha
    \left[
     \frac{L}{2\pi}\left(1+\frac{2d}L\right)
     -\frac12\,\dirac(\alpha)
    \right] f(\alpha).
 \end{eqnarray}

 The transition from the sum over $\mathbf{p}$ to the integral is performed
 according to Eq. (\ref{sumToIntPoP}).
 Then the density of states (\ref{DOS0}) is,
 \begin{eqnarray*}
    \rho(E)&=\frac{2S}{(2\pi)^2}\sum\limits_\alpha
             \int\!\!\dd\mathbf{p}\,\dirac\left(E-\frac{\hbar^2(p^2+\alpha^2)}{2m}\right)\\
           &=\frac{S}{\pi}  \sum\limits_\alpha
            \int\limits_0^\infty\!\!\dd{p}\,p\,\,\dirac\left(E-\frac{\hbar^2(p^2+\alpha^2)}{2m}\right)
           =\frac{S}{2\pi}\frac{2m}{\hbar^2} \sum\limits_\alpha\heav\left(\frac{2mE}{\hbar^2}-\alpha^2\right).
 \end{eqnarray*}
 Transformation from the sum over $\alpha$ to the integral
 according to Eq. (\ref{sumToIntPoAlfa}) leads to,
 \begin{eqnarray}\label{DOSIBM}
    \rho(E)&=\frac{S}{2\pi}\frac{2m}{\hbar^2}\!\!
    \int\limits_{0}^{+\infty}\!\!\!\dd \alpha\!
    \left[
     \frac{L}{2\pi}\!\left(\!1+\frac{2d}L\right)\!
     -\!\frac12\dirac(\alpha)
    \right]\! \heav\!\left(\!\frac{2mE}{\hbar^2}-\alpha^2\!\right)\!\nonumber\\
    &=\frac{SL}2\frac{\sqrt{2}m^{3/2}}{\pi^2\hbar^3}\sqrt{E}+
      S\left(\frac{\sqrt{2}m^{3/2}d}{\pi^2\hbar^3}\sqrt{E}
      -\frac{m}{4\pi\hbar^2}\right),
 \end{eqnarray}
 where the first term (which is proportional to the volume ${SL}$)
 is the extensive contribution, and the second term is the surface contribution to the density of states.
 This expression coincides with the expression for the density of states,
  obtained in Ref. \cite{Paash3}, if we put ${d=0}$ in Eq. (\ref{DOSIBM}).

 It should be noted that
 the density of states of non-interacting homogeneous system is,
 \[
  \rho_\unif(E)=V\frac{\sqrt{2}m^{3/2}}{\pi^2\hbar^3}\sqrt{E},
 \]
 which coincides with the first term of Eq.~(\ref{DOSIBM}).

\end{document}